\documentclass[fleqn,usenatbib]{mnras}

\usepackage{newtxtext,newtxmath}

\usepackage[T1]{fontenc}
\usepackage{ae,aecompl}

\usepackage{graphicx}	
\usepackage{amsmath}	
\usepackage{xcolor}
\usepackage{array}

\newcommand{\vect}[1]{\boldsymbol{#1}}
\newcommand{\rms}[1]{\text{rms}(#1)}
\newcommand{\T}[1]{T_{\text{#1}}}
\newcommand{\uK}{\text{K}}

\newcommand{\hT}[1]{\hat{T}_{\text{#1}}}
\newcommand{\hG}{\hat{G}}
\newcolumntype{x}[1]{>{\centering\let\newline\\\arraybackslash\hspace{0pt}}p{#1}}

\title[Effects of model incompleteness]{Effects of model incompleteness on the drift-scan calibration of radio telescopes}

\author[Gehlot et al.]{Bharat~K.~Gehlot$^{1}$\thanks{E-mail: kbharatgehlot@gmail.com (BKG)}, Daniel~C.~Jacobs$^{1}$,
Judd~D.~Bowman$^{1}$,
Nivedita~Mahesh$^{1}$,
\newauthor
Steven~G.~Murray$^{1}$,
Matthew~Kolopanis$^{1}$,
Adam~P.~Beardsley$^{25,1,\dagger}$,
Zara~Abdurashidova$^{2}$,
\newauthor
James~E.~Aguirre$^{3}$,
Paul~Alexander$^{4}$,
Zaki~S.~Ali$^{2}$,
Yanga~Balfour$^{5}$,
Gianni~Bernardi$^{6,7,5}$,
\newauthor
Tashalee~S.~Billings$^{3}$,
Richard~F.~Bradley$^{8}$,
Phil~Bull$^{9}$,
Jacob~Burba$^{10}$,
Steve~Carey$^{4}$,
\newauthor
Chris~L.~Carilli$^{11}$,
Carina~Cheng$^{2}$,
David~R.~DeBoer$^{2}$,
Matt~Dexter$^{2}$,
Eloy~de~Lera~Acedo$^{4}$,
\newauthor
Joshua~S.~Dillon$^{2,\dagger}$,
John~Ely$^{4}$,
Aaron~Ewall-Wice$^{12}$,
Nicolas~Fagnoni$^{4}$,
Randall~Fritz$^{5}$,
\newauthor
Steven~R.~Furlanetto$^{13}$,
Kingsley~Gale-Sides$^{4}$,
Brian~Glendenning$^{11}$,
Deepthi~Gorthi$^{2}$,
\newauthor
Bradley~Greig$^{14}$,
Jasper~Grobbelaar$^{5}$,
Ziyaad~Halday$^{5}$,
Bryna~J.~Hazelton$^{15,16}$,
\newauthor
Jacqueline~N.~Hewitt$^{12}$,
Jack~Hickish$^{2}$,
Austin~Julius$^{5}$,
Nicholas~S.~Kern$^{12}$,
Joshua~Kerrigan$^{10}$,
\newauthor
Piyanat~Kittiwisit$^{17}$,
Saul~A.~Kohn$^{3}$,
Adam~Lanman$^{10}$,
Paul~La~Plante$^{2,3}$,
Telalo~Lekalake$^{5}$,
\newauthor
David~Lewis$^{1}$,
Adrian~Liu$^{18}$,
Yin-Zhe~Ma$^{19}$,
David~MacMahon$^{2}$,
Lourence~Malan$^{5}$,
\newauthor
Cresshim~Malgas$^{5}$,
Matthys~Maree$^{5}$,
Zachary~E.~Martinot$^{3}$,
Eunice~Matsetela$^{5}$,
\newauthor
Andrei~Mesinger$^{20}$,
Mathakane~Molewa$^{5}$,
Raul~A.~Monsalve$^{18,1,24}$
Miguel~F.~Morales$^{15}$,
\newauthor
Tshegofalang~Mosiane$^{5}$,
Abraham~R.~Neben$^{12}$,
Bojan~Nikolic$^{4}$,
Aaron~R.~Parsons$^{2}$,
\newauthor
Robert~Pascua$^{2}$,
Nipanjana~Patra$^{2}$,
Samantha~Pieterse$^{5}$,
Jonathan~C.~Pober$^{10}$,
\newauthor
Nima~Razavi-Ghods$^{4}$,
Jon~Ringuette$^{15}$,
James~Robnett$^{11}$,
Kathryn~Rosie$^{5}$,
\newauthor
Mario~G.~Santos$^{5,21}$,
Peter~Sims$^{18}$,
Craig~Smith$^{5}$,
Angelo~Syce$^{5}$,
Max~Tegmark$^{12}$,
\newauthor
Nithyanandan~Thyagarajan$^{11,\ddagger}$,
Peter~K.~G.~Williams$^{22,23}$,
Haoxuan~Zheng$^{12}$
\\
$^{1}$ School of Earth and Space Exploration, Arizona State University, Tempe, AZ\\
$^{2}$ Department of Astronomy, University of California, Berkeley, CA\\
$^{3}$ Department of Physics and Astronomy, University of Pennsylvania, Philadelphia, PA\\
$^{4}$ Cavendish Astrophysics, University of Cambridge, Cambridge, UK\\
$^{5}$ South African Radio Astronomy Observatory, Black River Park, 2 Fir Street, Observatory, Cape Town, 7925, South Africa\\
$^{6}$ Department of Physics and Electronics, Rhodes University, PO Box 94, Grahamstown, 6140, South Africa\\
$^{7}$ INAF-Istituto di Radioastronomia, via Gobetti 101, 40129 Bologna, Italy\\
$^{8}$ National Radio Astronomy Observatory, Charlottesville, VA\\
$^{9}$ Queen Mary University London\\
$^{10}$ Department of Physics, Brown University, Providence, RI\\
$^{11}$ National Radio Astronomy Observatory, Socorro, NM\\
$^{12}$ Department of Physics and MIT Kavli Institute for Astrophysics and Space Research, MIT, Cambridge, MA\\
$^{13}$ Department of Physics and Astronomy, University of California, Los Angeles, CA\\
$^{14}$ School of Physics, University of Melbourne, Parkville, VIC 3010, Australia\\
$^{15}$ Department of Physics, University of Washington, Seattle, WA\\
$^{16}$ eScience Institute, University of Washington, Seattle, WA\\
$^{17}$ School of Chemistry and Physics, University of KwaZulu-Natal, Westville Campus, Private Bag X54001, Durban, South Africa\\
$^{18}$ Department of Physics and McGill Space Institute, McGill University, 3600 University Street, Montreal, QC H3A 2T8, Canada\\
$^{19}$ School of Chemistry and Physics, University of KwaZulu-Natal, Westville Campus, Durban, 4000, South Africa\\
$^{20}$ Scuola Normale Superiore, 56126 Pisa, PI, Italy\\
$^{21}$ Department of Physics and Astronomy,  University of Western Cape, Cape Town, 7535, South Africa\\
$^{22}$ Center for Astrophysics, Harvard \& Smithsonian, Cambridge, MA\\
$^{23}$ American Astronomical Society, Washington, DC\\
$^{24}$ Facultad de Ingenier\'{i}a, Universidad Cat\'{o}lica de la Sant\'{i}sima Concepci\'{o}n, Alonso de Ribera 2850, Concepci\'{o}n, Chile \\
$^{25}$ Department of Physics, Winona State University, Winona, MN \\
$^{\dagger}$ NSF Astronomy and Astrophysics Postdoctoral Fellow\\
$^{\ddagger}$ NRAO Jansky Fellow
}

\date{Accepted XXX. Received YYY; in original form ZZZ}

\pubyear{2020}

\begin{document}
\label{firstpage}
\pagerange{\pageref{firstpage}--\pageref{lastpage}}
\maketitle
\newpage
\begin{abstract}

Precision calibration poses challenges to experiments probing the redshifted 21-cm signal of neutral hydrogen from the Cosmic Dawn and Epoch of Reionization ($z\sim30-6$). In both interferometric and global signal experiments, systematic calibration is the leading source of error. Though many aspects of calibration have been studied, the overlap between the two types of instruments has received less attention. We investigate the sky based calibration of total power measurements with a HERA dish and an EDGES style antenna to understand the role of auto-correlations in the calibration of an interferometer and the role of sky in calibrating a total power instrument. Using simulations we study various scenarios such as time variable gain, incomplete sky calibration model, and primary beam model. We find that temporal gain drifts, sky model incompleteness, and beam inaccuracies cause biases in the receiver gain amplitude and the receiver temperature estimates. In some cases, these biases mix spectral structure between beam and sky resulting in spectrally variable gain errors. Applying the calibration method to the HERA and EDGES data, we find good agreement with calibration via the more standard methods. Although instrumental gains are consistent with beam and sky errors similar in scale to those simulated, the receiver temperatures show significant deviations from expected values. While we show that it is possible to partially mitigate biases due to model inaccuracies by incorporating a time-dependent gain model in calibration, the resulting errors on calibration products are larger and more correlated. Completely addressing these biases will require more accurate sky and primary beam models.
\end{abstract}

\begin{keywords}
dark ages, reionization, first stars -- methods: statistical -- methods: data analysis -- techniques: interferometric -- instrumentation: miscellaneous -- instrumentation: interferometers 
\end{keywords}



\section{Introduction}

Observations of the highly redshifted 21-cm signal of neutral hydrogen (HI) from the Cosmic Dawn ($z\sim 30-12$) and the Epoch of Reionization ($z\sim 12-6$) have the potential to uncover a wealth of information about the properties of the first luminous objects (e.g. first stars and galaxies), intergalactic medium as well as fundamental physics questions. This promising avenue has motivated the development of instruments targeting the low-frequency band. These arrays are both interferometric arrays like e.g. the Giant Meterwave Radio Telescope (GMRT; \citealt{paciga2011}), the Low Frequency Array (LOFAR; \citealt{vanhaarlem2013}), the Murchison Widefield Array (MWA; \citealt{tingay2013,bowman2013}), the Precision Array to Probe Epoch of Reionization (PAPER; now decommissioned; \citealt{parsons2010}), the Hydrogen Epoch of Reionization Array (HERA; \citealt{deboer2017}), the Owens Valley Long Wavelength Array (OVRO-LWA; \citealt{eastwood2018,eastwood2019}), the New Extension in Nan\c cay Upgrading loFAR (NENUFAR; \citealt{zarka2012}), and the upcoming Square Kilometer Array (SKA; \citealt{mellema2013,koopmans2015}),  as well as single-receiver radiometers including the Experiment to Detect the Global Epoch of reionization Signature (EDGES; \citealt{bowman2018}), the Shaped Antenna measurement of the background RAdio Spectrum-2 (SARAS2; \citealt{singh2017}), the Large-aperture Experiment to Detect the Dark Ages (LEDA; \citealt{bernardi2016}), the Probing Radio Intensity at high $z$ from Marion (PRIZM; \citealt{philip2019}). All these experiments are working towards measuring the brightness temperature fluctuations of the redshifted 21-cm HI signal and the sky-averaged 21-cm signal (or global 21-cm signal) from the epochs of Cosmic Dawn and Reionization. 

The redshifted 21-cm signal (both global signal and fluctuations) is extremely faint. It is contaminated by bright astrophysical foregrounds (Galactic diffuse and free-free emission, supernova remnants, radio galaxies and clusters etc.) that are several orders of magnitude brighter than the signal of interest, ionosphere of the Earth, and instrumental imperfections, e.g. direction independent and dependent instrumental response, frequency-dependent instrumental bandpass, polarization leakage etc. These contaminations make extraction of the 21-cm signal from the observed signal an extremely challenging process.

Calibration of instruments used by 21-cm cosmology experiments is a daunting task and needs to be performed with great accuracy and precision (with an error level $\sim 10^{-5}$) to achieve a dynamic range high enough to detect the faint 21-cm signal. Most interferometric 21-cm cosmology experiments use calibration methods that rely on knowledge of the sky and/or array layout (redundancy between baselines). These methods utilize cross-correlation products to obtain per antenna complex gain (both direction independent and dependent) that are used to correct the observed cross-correlations \citep{mitchell2008,salvini2014,yatawatta2015,li2018}. On the other hand, global 21-cm signal experiments with single-element radiometers use calibration methods that require switching between various loads (see e.g. \citealt{pauliny-toth1962,rogers2012,monsalve2017}) and absolute receiver temperature measurements by putting antenna+receiver in anechoic chambers of known temperatures (only possible for miniature antennas, see e.g. \citealt{an1993}). However, many experiments have also explored the use of total power sky measurements (or auto-correlations) for various types of calibration, such as bandpass amplitude, signal chain reflection and mutual coupling calibration (see e.g. \citealt{rogers2004,ewall-wice2016,monsalve2017,singh2018,mozdzen2019,barry2019,li2019,kern2019, monsalve2021}, HERA memos for various implementations). Various statistical estimators used in 21-cm cosmology interferometric experiments also require correction of bias introduced due to instrumental noise temperature and correct propagation of thermal uncertainties in the analyses (see e.g. \citealt{trott2016,kolopanis2019,mertens2020}). The very high accuracy and precision required in this measurement has led to the exploration of auto-correlation (or total power) based calibration by both types of experiments.

Most methods to calibrate instruments using auto-correlations/total power measurements employ known sky brightness maps and primary beam patterns to obtain calibration products to correct the observed data. \cite{rogers2004} describes one such method that utilizes sky-brightness maps (or their simulations) to calibrate antenna arrays by obtaining gain amplitudes (bandpass) and receiver noise temperature per frequency channel. This method uses measured power or auto-correlations (in case of single element or interferometers, respectively) and known beam averaged sky-brightness temperature as a function of sidereal time and frequency to obtain receiver temperature and bandpass gain amplitude. However, such methods are susceptible to errors and biases due to various factors, such as instrumental instability in time, sky model incompleteness and inaccuracies in the primary beam model used for calibration. Recently, \cite{li2021} investigated gain, sky temperature, and receiver temperature variations in the MeerKAT receiver system  using the correlated (1/f) noise analyses of the South Celestial pole tracking observations with MeerKAT. 

In this paper, we use simulations of auto-correlations (total power) to investigate various factors that could produce bias in calibration and possible ways to mitigate these biases. We further use delay spectrum analysis to study the effect of model incompleteness on calibration products from auto-correlation based calibration. The paper is organized as follows: section \ref{sec:autocalibration} defines the auto-correlation based interferometric calibration method. Section \ref{sec:methodology} provides a brief description of the instrument models and the auto-correlation simulations we have used for analysis. We discuss various model incompleteness effects in section \ref{sec:model-incompleteness}, and provide a comparison of these effects in delay space in section~\ref{sec:delay-spec}. In section~\ref{sec:real-data-calibration}, we discuss the calibration of HERA and EDGES data using the auto-correlation based calibration. Finally, we summarize our tests and provide some context discussion in section \ref{sec:summary}.

\section{Instrumental calibration using auto-correlations/total power measurements}\label{sec:autocalibration}

Interferometers correlate every signal from one antenna element with itself (auto-correlations) and signals from other antenna elements (cross-correlations).  Typically, in arrays like the JVLA, which are not equipped with Dicke switching radiometers, auto-correlations are usually not employed in calibration. However, responding to the challenge of calibrating on wide fields without signal loss/suppression \citep{patil2016,barry2016}, 21-cm experiments have used auto-correlations to calibrate signal chain reflections \citep{barry2019,li2019,kern2019,kern2020a} and to make an independent measure of absolute calibration (see e.g. \href{http://reionization.org/wp-content/uploads/2013/03/HERA_34_Modeling_Sky_Temperature.pdf}{HERA memo \#34}, \citealt{bowman2007}). The latter is the focus of our investigation here.

For a stable, linear system, auto-correlations (or total power) measured by an antenna element ($R_i$) can be modelled as the sum of the temperature due to sky power ($\T{sky}$) and the receiver noise temperature ($\T{rxr}$) attenuated by the receiver gain ($g_i$),
\begin{equation}\label{eqn:autocorr}
\begin{split}
R_i (\nu,t) & = g_i(\nu) g_i^*(\nu)(\T{sky} (\nu, t) + \T{rxr}) \\
& = G_i(\nu) (\T{sky}(\nu,t) + \T{rxr}) \, .    
\end{split}
\end{equation}

Note that $g_i$ and $\T{rxr}$ are assumed to be constant in time. For convenience, we use the notation $G_i = |g_i|^2$ hereafter. Though it is standard practice to describe receiver temperature in terms of equivalent sky power thus making it dependent on gain, in reality, gain and noise are not simply related physical properties. Here, we redefine auto-correlations $R$ in terms of internal noise $n$, resulting in the following equation,  
\begin{equation}\label{eqn:red-autocorr}
R_i (\nu,t) = G_i(\nu)\T{sky}(\nu, t) + n_i(\nu)\,,
\end{equation}
where $n_i$ is the additional noise bias (referred to as noise figure hereafter) due to the antenna element and can be defined in terms of receiver temperature $\T{rxr} = n/G$. The power due to the sky in temperature units $\T{sky}(\nu, t)$ can be modelled with a sky brightness temperature map $\T{map}$ (see e.g. \cite{haslam1982,deoliveiracosta2008,zheng2017}) which is a function of apparent coordinates $(\theta,\phi)$ at time $t$, and frequency $\nu$. Expected sky temperature spectrum $\T{sky}(\nu,t)$ may be calculated as a weighted average of $\T{map}$ with weights determined by the antenna primary beam $A(\theta,\phi,\nu)$:
\begin{equation}\label{eqn:Tsky}
\T{sky}(\nu,t) = \dfrac{\int_{\Omega} A(\theta,\phi,\nu) T_{\text{map}}(\theta,\phi,\nu,t) d\Omega}{\int_{\Omega} A(\theta,\phi,\nu) d\Omega}\, .    
\end{equation}

\begin{figure*}
\centering
\includegraphics[width=0.9\textwidth]{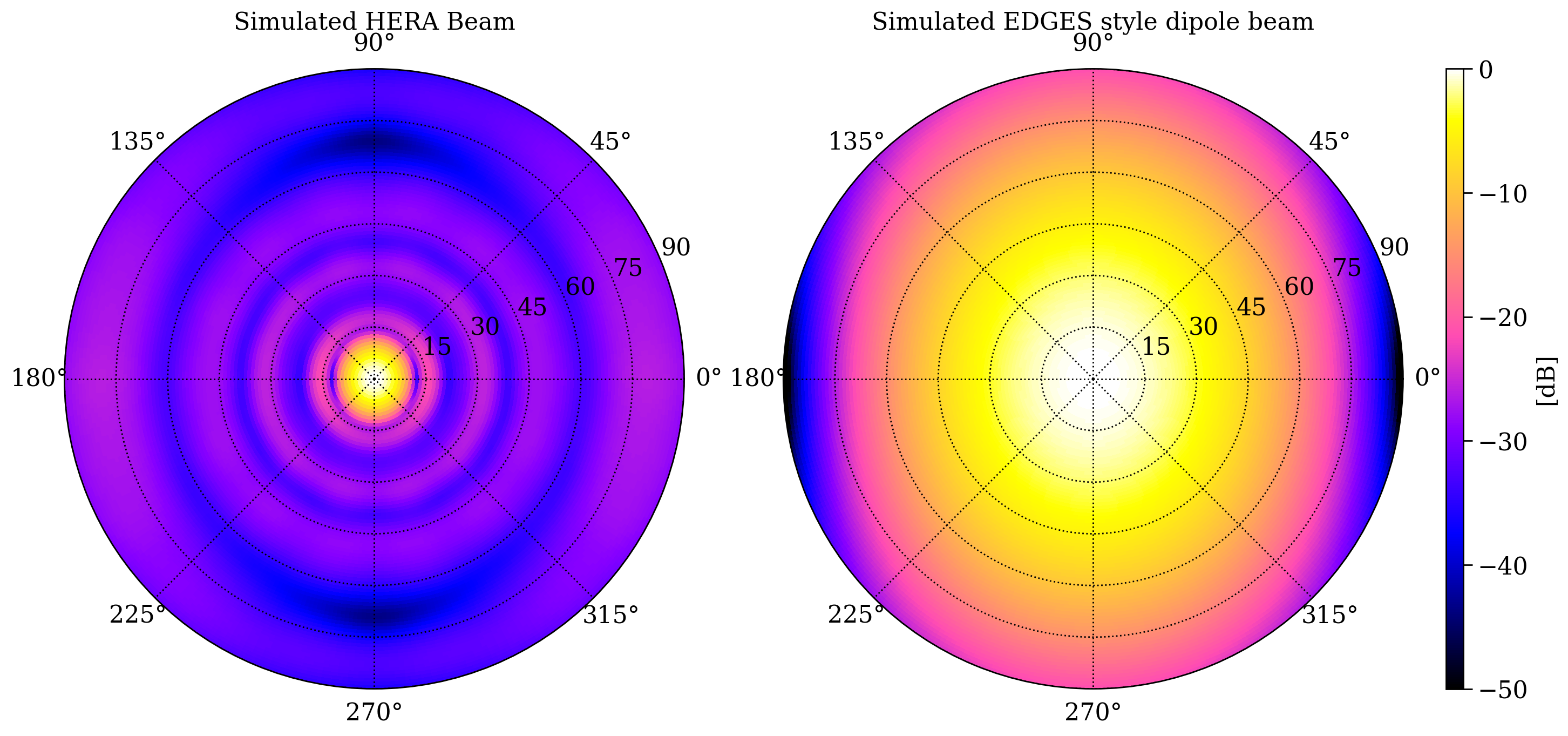}
    \caption{Simulated primary beam power patterns for the two types of receiver designs used in the analysis. Left panel: Simulated primary beam power pattern (single polarization) of a HERA dish with the dipole feed (PAPER type). Right panel: Simulated primary beam power pattern for a EDGES style dipole (Gaussian-type). Beam patterns are shown in polar projection where the spokes represent azimuth angle, and the dotted circles represent the zenith angle with $15^{\circ}$ separation between consecutive circles.} 
\label{fig:HERA-Beam}
\end{figure*}

Auto-correlations measured by an interferometer may be used to determine the antenna gain amplitude $|g_i|$ and the receiver temperature $\T{rxr}$ if the primary beam of antenna element ($A$) and the brightness temperature of the sky ($\T{map}$) observed by the interferometer is well known. For a given antenna element, equation~\ref{eqn:red-autocorr} can be modelled as:
\begin{equation}\label{eqn:signal-model}
\vect{y} = \textbf{A}\vect{x} + \vect{\epsilon} \, , \\
\end{equation}
where
\begin{equation}\label{eqn:signal-matrices}
\begin{split}    
& \vect{y} = \begin{bmatrix} R[0] & ... & R[t] \end{bmatrix}^{T} \, , \\
& \vect{x} = \begin{bmatrix} G & n \end{bmatrix}^{T} \, , \\
& \textbf{A} = \begin{bmatrix} \T{sky}[0] & ... & \T{sky}[t] \\
                                1 & ... & 1 \end{bmatrix}^{T} \, ,
\end{split}
\end{equation}
and $\vect{\epsilon}$ is the noise vector. The least squares fit to equation~\ref{eqn:signal-model} can be written as 
\begin{equation}\label{eqn:best-fit}
\hat{\vect{x}} = (\mathbf{A}^T \mathbf{C}^{-1} \mathbf{A})^{-1}\mathbf{A}^T \mathbf{C}^{-1} \vect{y} \, ,\\ 
\end{equation}
where $\hat{\vect{x}}$ is a vector consisting of best fit gain ($\hG$) and noise figure ($\hat{n}$). The best fit receiver temperature ($\hT{rxr}$) can then be obtained using the following relation:
\begin{equation}
 \hT{rxr} = \dfrac{\hat{n}}{\hG}\, .   
\end{equation}
Uncertainties on best-fit parameters can then be obtained from the covariance of estimated parameters $\textbf{C}_{\hat{\vect{x}}}$ given by

\begin{equation}\label{eqn:best-fit-error}
\begin{split}
& \textbf{C}_{\hat{\vect{x}}} = (\mathbf{A}^T \textbf{C}^{-1} \mathbf{A})^{-1}\, ,\\
& \text{Var}(\hG) = \hat{\sigma}_{00}^2\, , \\
& \text{Var}(\hT{rxr}) = \hT{rxr}^2 \sum_{i,j} (-1)^{i+j} \dfrac{\hat{\sigma}_{ij}^2}{\hat{x}_i \hat{x}_j}
\end{split}
\end{equation}

where $\mathbf{C}$ is covariance of data and $\hat{x}_{i}$ and $\hat{\sigma}_{ij}^2$ are elements of $\hat{\vect{x}}$ vector and $\textbf{C}_{\hat{\vect{x}}}$ matrix, respectively. Auto-correlations for every frequency channel can be independently calibrated to obtain $\hG$ and $\hT{rxr}$. The method described above assumes that receiver temperature and antenna gain do not vary over time, the primary beam is accurately known, and the sky model used in the calibration process is accurate. However, several uncertainties in the model, e.g. incomplete knowledge of sky brightness, incorrect beam model, temporal and spectral variations of antenna gains, instrumental effects such as cable reflections and cross-talk between antennas can cause errors and bias in the estimation of $G_i$ and $\T{rxr}$ from auto-correlations. Here, our aim is to quantify the effect of antenna gain variations, incomplete sky model and incorrect beam model on auto-correlation based calibration using simulated auto-correlations of HERA dish and dipole type receivers.

\begin{figure*}
\centering
\includegraphics[width=\textwidth]{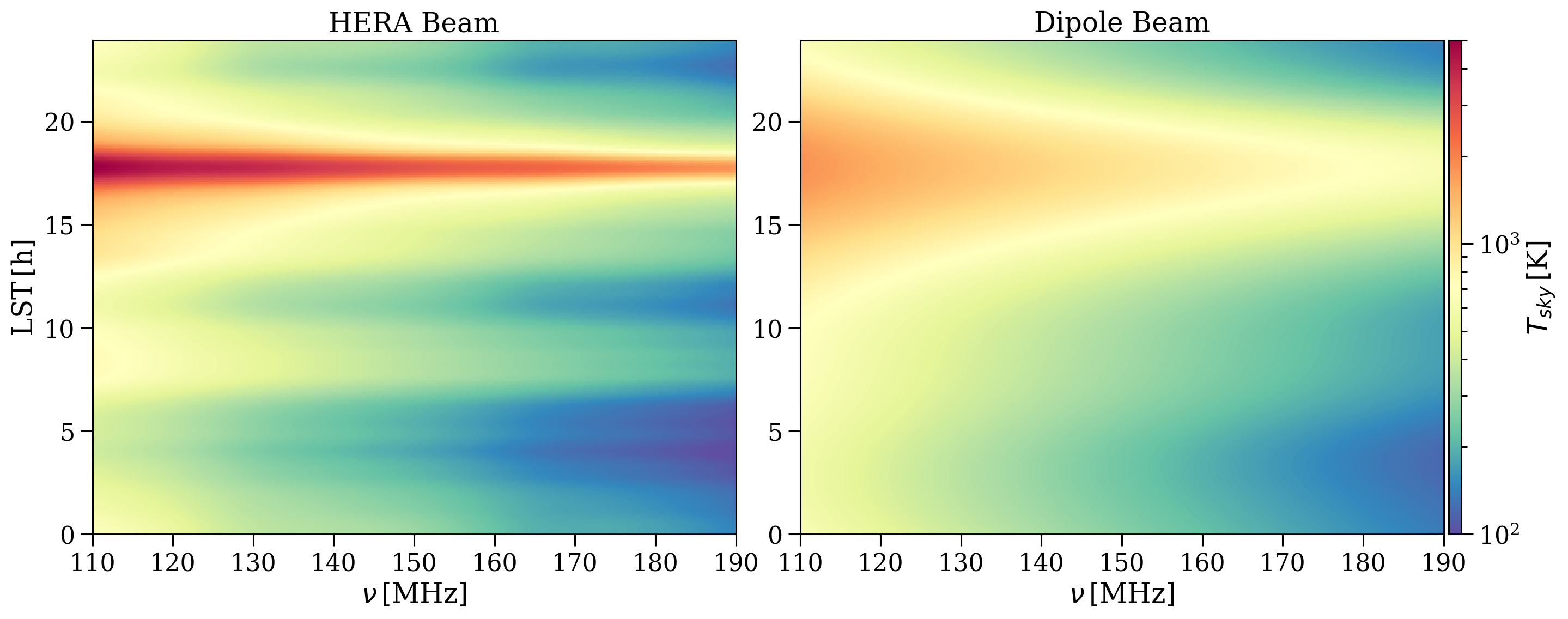}
    \caption{Here we show the primary beam averaged sky temperature profiles for the HERA dish (left panel) and dipole type (right panel) receivers as a function of LST and frequency. These profiles were calculated using the GSM2008 sky model of diffuse emission. Large $\T{sky}$ value around LST = 17.5~h corresponds to the Galactic center transiting through the zenith. Because of the narrower primary beam, the sky temperature profile for the HERA dish shows more structure (and stronger peak for the Galactic center transit) along the LST axis compared to the dipole antenna with the wider field of view.} 
\label{fig:Tsky}
\end{figure*} 

\section{Methodology}\label{sec:methodology}

\subsection{Instrument Models}\label{subsec:instrument}
Here we consider two element types, a drift-scan dish representing HERA and a dipole representing EDGES. In both cases, we make simplifying approximations to the instrument model which allows to control the variations. These approximations are reasonable and generally used in simulations without causing any significant deviations from real instruments.

HERA is a next-generation radio interferometer located in the Karoo desert, South Africa ($30.7224^{\circ}$S, $21.4278^{\circ}$E, 1100~m elevation). It is designed to measure the redshifted 21-cm signal of neutral hydrogen from the Cosmic Dawn and Epoch of Reionization (z = 25 to 6). It is a densely packed drift-scan array of parabolic dishes of 14~m diameter and is currently in the build-out phase. HERA baseline design is highly redundant, with 320 dishes closely packed into a hexagonal 300\,m-wide grid and 30 outrigger dishes providing baselines up to 3 km. In the first iteration, HERA used feeds based on PAPER antenna design (dipole feeds) operating in the frequency range of 100-200~MHz (100~MHz operational bandwidth). The final design will have broadband `Vivaldi' feeds with the operational frequency bandwidth of 200 MHz (50-250MHz) to cover both Epoch of Reionization and Cosmic Dawn frequencies. Readers are referred to \cite{deboer2017} for detailed information about the HERA telescope. The first iteration with the dipole feed was the subject of several performance studies \citep{Patra2018, thyagarajan2016,ewall-wice2017} and used for a deep integration (see e.g. \citealt{kern2019,kern2020a,kern2020b} for HERA Phase-I calibration for 21-cm analyses). The HERA dipole-feed was given detailed studies with simulations by \cite{fagnoni2021}, who produced the beam model that we use here.\footnote{Beam model files are available at github.com/HERA-Team/HERA-Beams}

As a comparison point, we also include an isolated dipole antenna in our analysis. Though not specifically modelled on a single global experiment, the selected dipole antenna most closely aligns to the EDGES style broadband dipole. For the dipole receiver, we use a Gaussian type primary beam described as
\begin{equation}\label{eqn:dipole-beam}
    A(\theta) = \dfrac{1}{\sqrt{\pi}\theta_0} \exp{\left[- \left(\dfrac{\theta}{\theta_0}\right)^2 \right]}\, ,
\end{equation}
where $\theta$ is the co-latitude (such that $\theta = 0$ at zenith), and  $\theta_0$ is calculated from the Full-Width Half Maximum (FWHM) that varies with frequency as
\begin{equation}
    \text{FWHM} = 72^{\circ} \left( \dfrac{\nu}{140~\text{MHz}} \right)^{-1}. 
\end{equation}
The FWHM of the dipole beam is chosen to be $72^{\circ}$ at 140~MHz, which is equivalent to the FWHM of the EDGES high band antenna at the same frequency. Additionally, we use an elliptical azimuthal profile for the dipole antenna beam such that the final primary beam pattern becomes
\begin{equation}
    A_{\text{dipole}}(\theta,\phi) = A(\theta) (\cos^2\phi + \sin^2\phi \cos\theta)^2 \, .
\end{equation}
The final power pattern approximately matches the EDGES antenna primary beam (see e.g. \citealt{mahesh2021}). Figure \ref{fig:HERA-Beam} shows the primary beam power patterns of a single polarization of a HERA dish (with PAPER feed) and the Gaussian dipole. We use these two beam models to simulate auto-correlations for further analyses. Hereafter, the simulated primary beam for the HERA dish will be referred to as the HERA beam, and the simulated EDGES style dipole beam will be referred to as the dipole beam in figures and text.

\subsection{Simulating mock auto-correlations}\label{subsec:auto-simulation}

We produce mock auto-correlations using the following template,
\begin{equation}\label{eqn:simulation-template}
    R[\nu,t] = G(\T{sky}[\nu,t] + \T{rxr}) + \epsilon.
\end{equation}
We choose a receiver temperature $\T{rxr} = 150\,\text{K}$ for both antenna designs and simulate mock auto-correlations for the frequency range $\nu = 110-190\,$MHz. The above equation also requires averaged sky temperature $\T{sky}[\nu,t]$ as a function of time and frequency. We use equation~\ref{eqn:Tsky} to calculate spatially averaged sky temperature in the 110-190 MHz frequency range. We use the Global Sky Model (GSM) of diffuse radio emission presented in \cite{deoliveiracosta2008} (GSM2008 hereafter) to obtain spatially averaged sky temperatures weighted with the HERA and the dipole beams for 0-24~hours LST range. Figure \ref{fig:Tsky} shows averaged sky temperature profiles obtained from GSM2008 for both HERA and dipole beams. The selection of $G$ values to simulate mock auto-correlations is dependent on model incompleteness scenarios we have investigated. For temporal gain variation scenarios, the gain is time-dependent and the gain profile $G[t]$ is set such that it varies around $G_{0} = 1$. For other model incompleteness scenarios, the gain is assumed to be constant with value $G = 1$. The mock auto-correlations are sampled at time and frequency intervals of 5 minutes and 250\,kHz, respectively. The uncertainty on auto-correlations ($\sigma_R$) is proportional to the antenna temperature $T_{A}$ (sum of the beam-averaged sky temperature $\T{sky}$ and the receiver temperature $\T{rxr}$) and is given by the standard radiometer equation \citep{wilson2009}:
\begin{equation}\label{eqn:radiometer}
    \sigma_{R} = \dfrac{T_{A}}{\sqrt{\Delta \nu \Delta t}} = \dfrac{\T{sky}+\T{rxr}}{\sqrt{\Delta \nu \Delta t}}
\end{equation}
where $\Delta \nu$ and $\Delta t$ are integration time and frequency of the instrument and chosen to be $\Delta \nu = 5$~kHz and $\Delta t=1$~s for both antenna designs. For a given time and frequency, the noise ($\epsilon$) on each auto-correlation value is drawn from Gaussian distributions $\mathcal{N}(0,\sigma_R^2)$ with variance given by equation~\ref{eqn:radiometer}.

\begin{figure*}
\centering
\includegraphics[width=\textwidth]{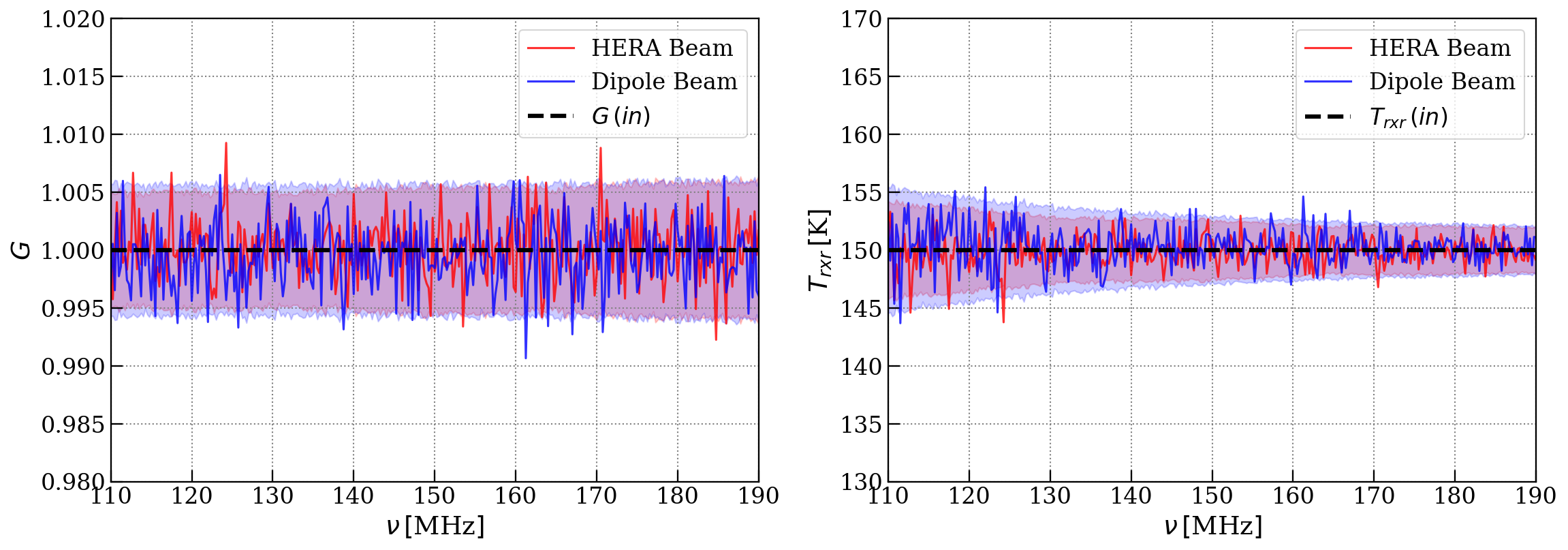}
    \caption{Calibration parameters for the reference simulation. Left panel:  $\hG$ as a function of frequency for HERA (red curve) and dipole (blue curve) primary beams. Right panel: same as the left panel but for $\hT{rxr}$. Black dashed lines show input parameter values for auto-correlations simulation. Shaded regions represent the $2\sigma$ error for corresponding parameters. Note that $2\sigma$ errors are placed around the input parameters for a clear representation of error levels. The calibration products fluctuate around input values in simulations with small errors and do not exhibit any visible bias.} 
\label{fig:reference_sim}  
\end{figure*} 

\section{Effect of Model Incompleteness}\label{sec:model-incompleteness}

In this section, we explore different effects that may cause errors and bias in $G$ and $\T{rxr}$ estimation. To set a reference, we produce auto-correlations using equation~\ref{eqn:simulation-template}, with default values for $G$, $\T{rxr}$ and fit for the parameters as described in section \ref{sec:autocalibration} using the same sky temperature model $\T{sky}$ used to simulate mock-auto-correlations. The results are shown in figure \ref{fig:reference_sim}. In an ideal scenario, where the sky temperature and beam model are perfectly known and antenna gain do not change with time and frequency, the calibration parameters $\hT{rxr}$ and $\hG$ are obtained with small rms error ($\sim0.2\%$ on $\hG$ and $\sim 1\%$ on $\hT{rxr}$) for both antenna designs i.e. $\rms{G} \sim 0.002$, $\text{rms}(\T{rxr})\sim 1.35\,\uK$ for the HERA dish, and $\rms{G}\sim 0.002$, $\text{rms}(\T{rxr})\sim 1.64\,\uK$ for the dipole antenna. This error is dominated by the uncertainty on mock auto-correlations. We observe that $\hT{rxr}$ has larger errors at lower frequencies than at higher frequencies due to the spectral dependence of the sky temperature. This spectral dependence affects the error on the additive term in the fitting process leading to larger uncertainty on $\hT{rxr}$ at the lower end of the frequency band. Moreover, errors on $\hG$ and $\T{rxr}$ for the dipole are larger than for the HERA dish at lower frequencies because $\T{sky}$ for the dipole antenna has less information (independent sky measurements) along the LST direction compared to the HERA dish. In the following sections, we investigate the effect of various types of model incompleteness such as antenna gain variation, sky model incompleteness and primary beam errors on auto-correlation based calibration. 

\begin{figure}
\centering
\includegraphics[width=\columnwidth]{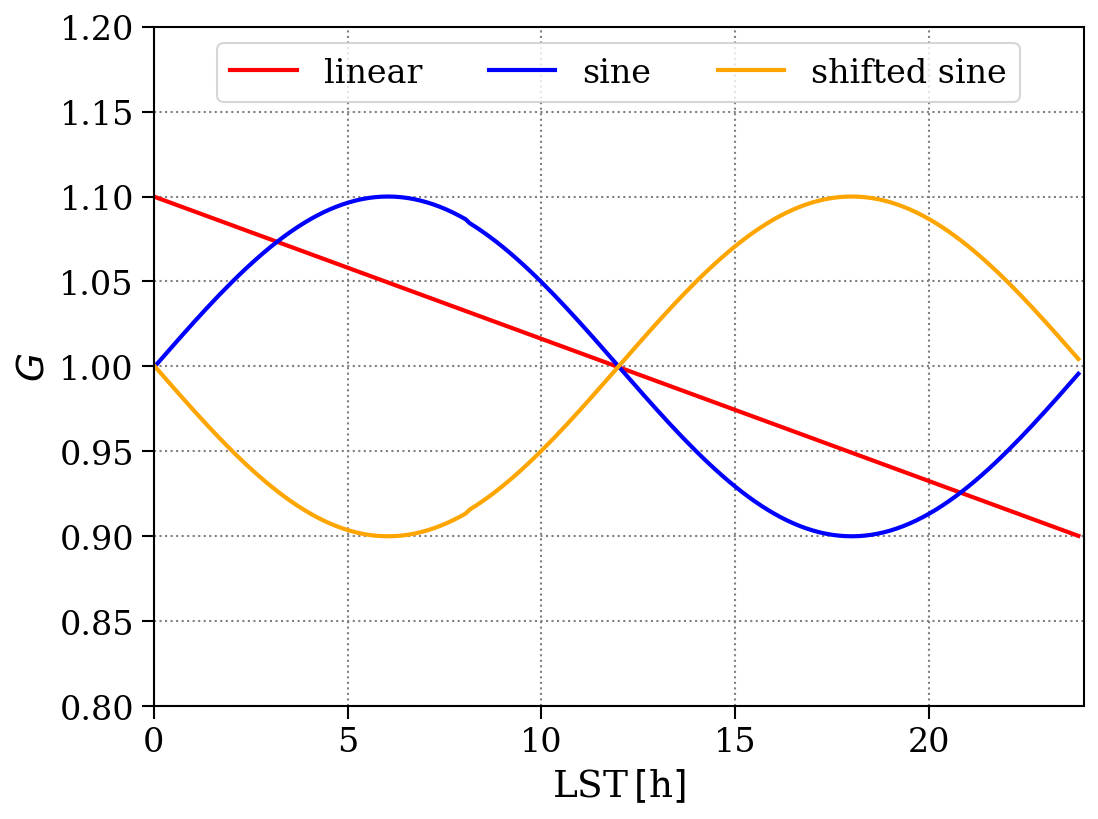}
    \caption{Gain profiles $G(t)$ for three types of gain variations with LST used in  section \ref{subsec:gain_variation} to simulate mock auto-correlations. Note that we assume that the instrument has a flat bandpass and the gains only vary in time.} 
\label{fig:Input_G_t}
\end{figure}

\subsection{Effect of antenna gain variation}\label{subsec:gain_variation}

The method described in section \ref{sec:autocalibration} assumes an ideal instrument that has stable antenna gains which do not vary with time throughout the observation and has a flat frequency response (bandpass). However, in reality, instruments impart a spectral structure on to the incoming sky signal. They can also have temporally varying antenna gains due to several factors such as ambient temperature variations and unstable electronics. In this section, we study the effect of temporal gain variation on the estimation of $\hT{rxr}$, and $\hG$ from auto-correlations. 

To quantify the effect of temporal variation of antenna gain, we introduce time dependent gain $G[t]$ in the simulation of mock auto-correlations instead of using a constant $G$ throughout the LST range. For simplification of the analysis, we consider three types of temporal gain variations represented by simple functions: (a) linear variation with LST, (b) sinusoidal variation that correlates with sky brightness, (c) sinusoidal variation that anti-correlates with sky brightness. Corresponding gain profiles are given by:

\begin{equation}\label{eqn:gain-time-variation}
\begin{split}
& G_a[t] = G_{0} - 0.1 \left(\dfrac{t}{12} - 1\right) \\
& G_b[t] = G_{0} + 0.1 \sin{\left(\dfrac{2 \pi t}{24}\right)} \\
& G_c[t] = G_{0} + 0.1 \sin{\left(2\pi - \dfrac{2\pi t}{24}\right)} \\
\end{split}
\end{equation}

where $G_{0} = 1$, and $0\leq t <   24$. The above equations are defined such that $G[t]$ varies around $G_{0}$ and $0.9 \leq G \leq 1.1$ for all three cases. Figure \ref{fig:Input_G_t} shows corresponding gain profiles. Gain drifts caused by ambient temperature variations over 24 hours are expected to be sinusoidal (profiles (b) or (c)). In observations that span shorter LST ranges of the order of a few hours, gain drifts are only a part of the sinusoidal function. These gain drifts can be approximated by type (a) variation. For example, in \cite{kern2020b}, it was reported that average gains for HERA antennas drift by $\sim5-6\%$ over 6-hour range. The drift is approximately linear in time and seems to be anti-correlated with the ambient temperature (as reported by a weather station nearby).  Additionally, gain drifts due to diurnal temperature variations are expected to follow similar behavior as the latter two cases. 

\begin{figure*}
\centering
\includegraphics[width=\textwidth]{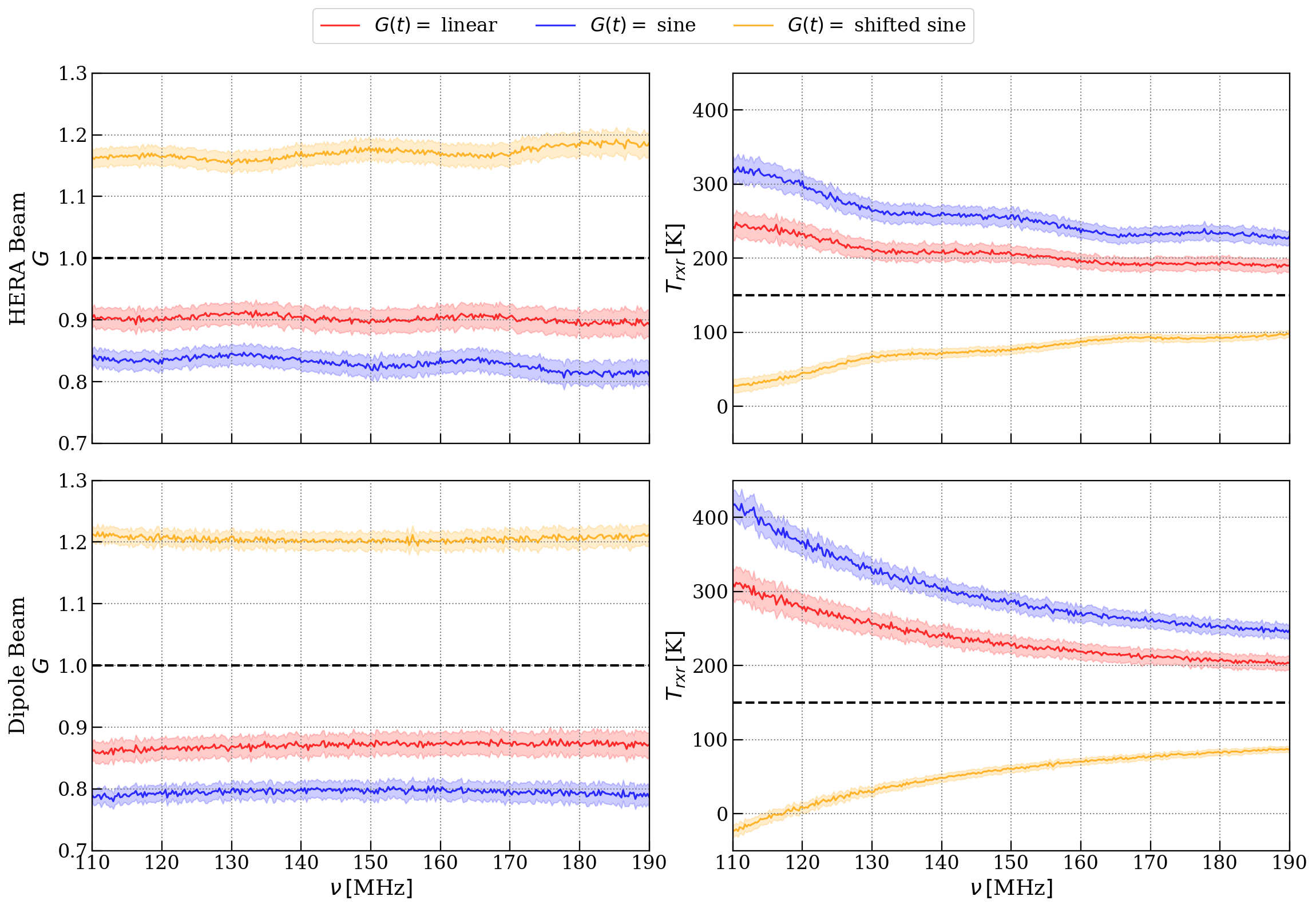}
    \caption{Calibration products for simulations with time-dependent gain $G(t)$ (three types of gain variations) for the two antenna designs (top and bottom rows). Left column: $\hG$ obtained from the calibration as a function of frequency. Right panel: same as left but for $\hT{rxr}$. Dashed lines in the left column correspond to the time-averaged profile of input gain ($\langle G \rangle_t$). Calibration products are biased for all three types of gain variations and corresponding uncertainties are smaller than the bias for both $\hG$ and $\hT{rxr}$.} 
\label{fig:time_dependence}  
\end{figure*} 

We use the above-described gain profiles to generate mock auto-correlations and calibrate those using the GSM2008 sky model to obtain a single $\hG$ and $\hT{rxr}$ value (per frequency channel) for 24-hours of simulated auto-correlations such that complete behavior of gain drifts is captured. The parameters obtained for different gain variation scenarios are shown in figure \ref{fig:time_dependence}. Because mock auto-correlations have a time-dependent gain, the obtained parameters are biased. The bias level is different for different types of gain drift and is also dependent on antenna design with larger bias for antennas with wider field of view. 

We note that the bias in $\hG$ for the HERA beam (top left in \ref{fig:time_dependence}) shows a weak spectral ripple (peak to peak variation of $\sim1$~percent) for all three gain variation scenarios. This is caused by the spectral variation in the HERA primary beam coupling into the model. The auto-correlations for a given frequency channel are brightest when the galactic center is in the field of view. During these LSTs, the sky model displays a spectral ripple on a 30MHz period. Under the proposed theory, gain varies with time causing times with the ripple to receive slightly more weight. Tests excluding the Galactic Center significantly reduce the ripple. Comparing the gain variation profiles in figure~\ref{fig:Input_G_t} we see that $\hG$ that gain amplitude correlates or anti-correlates with galactic center LSTs. The $\hT{rxr}$ values show an opposite trend because of the inverse dependence on gain. This effect is exclusive to the HERA beam which uses a fully frequency dependent EM simulation. $\hG$ for the dipole beam, which uses a Gaussian, is approximately constant. In addition to this, the bias in $\hT{rxr}$ for both antenna designs shows an increase in estimated values as frequency decreases. Spectral dependence of the sky temperature is likely the cause of this increase, leading to larger residuals (hence larger bias) at lower frequencies. This suggests that temporal gain drifts of $\sim 10\%$ over 24 hours can introduce significant bias in both $\hG$ and $\hT{rxr}$ that is also dependent on the primary beam passband. Antenna designs with a narrower field of view show relatively lower bias in $\hT{rxr}$ (approximately similar bias levels in $\hG$).

For the HERA antenna, the rms uncertainty on $\hG$ with respect to its rms amplitude is similar for linear and sinusoidal type gain variations and relatively lower for sinusoidal variations that anti-correlate with the Galactic Center LST with corresponding levels $\text{rms}(G_{\text{a}})\sim 1.05\%$, $\text{rms}(G_{\text{b}})\sim 1.03\%$ and $\text{rms}(G_{\text{c}})\sim 0.73\%$ respectively. The rms uncertainty in $\hT{rxr}$ however are similar for linear and sinusoidal type variations with corresponding rms levels of $\text{rms}(T_{\text{a}})\sim 2.8\%$ and $\text{rms}(T_{\text{b}})\sim 2.4\%$ but is relatively larger for variations that anti-correlate with the Galactic Center with rms level of $\text{rms}(T_{\text{c}})\sim 4.05\%$. The uncertainties on the calibration parameters for the dipole antenna are slightly smaller than for the HERA antenna but show a similar behavior. The rms uncertainty (with respect to the rms amplitude) on $\hG$ and $\hT{rxr}$ is $\text{rms}(G)\sim 1.04\%, \ 0.92\%, \ 0.61\%$ and $\text{rms}(T)\sim 2.8\%, \ 2.2\%, \ 4.7\%$ for (a), (b) and (c) types of gain variations, respectively. Note that the uncertainties for both antenna designs are smaller than the bias in all three cases of gain variation.
 
 \begin{figure*}
\centering
\includegraphics[width=\textwidth]{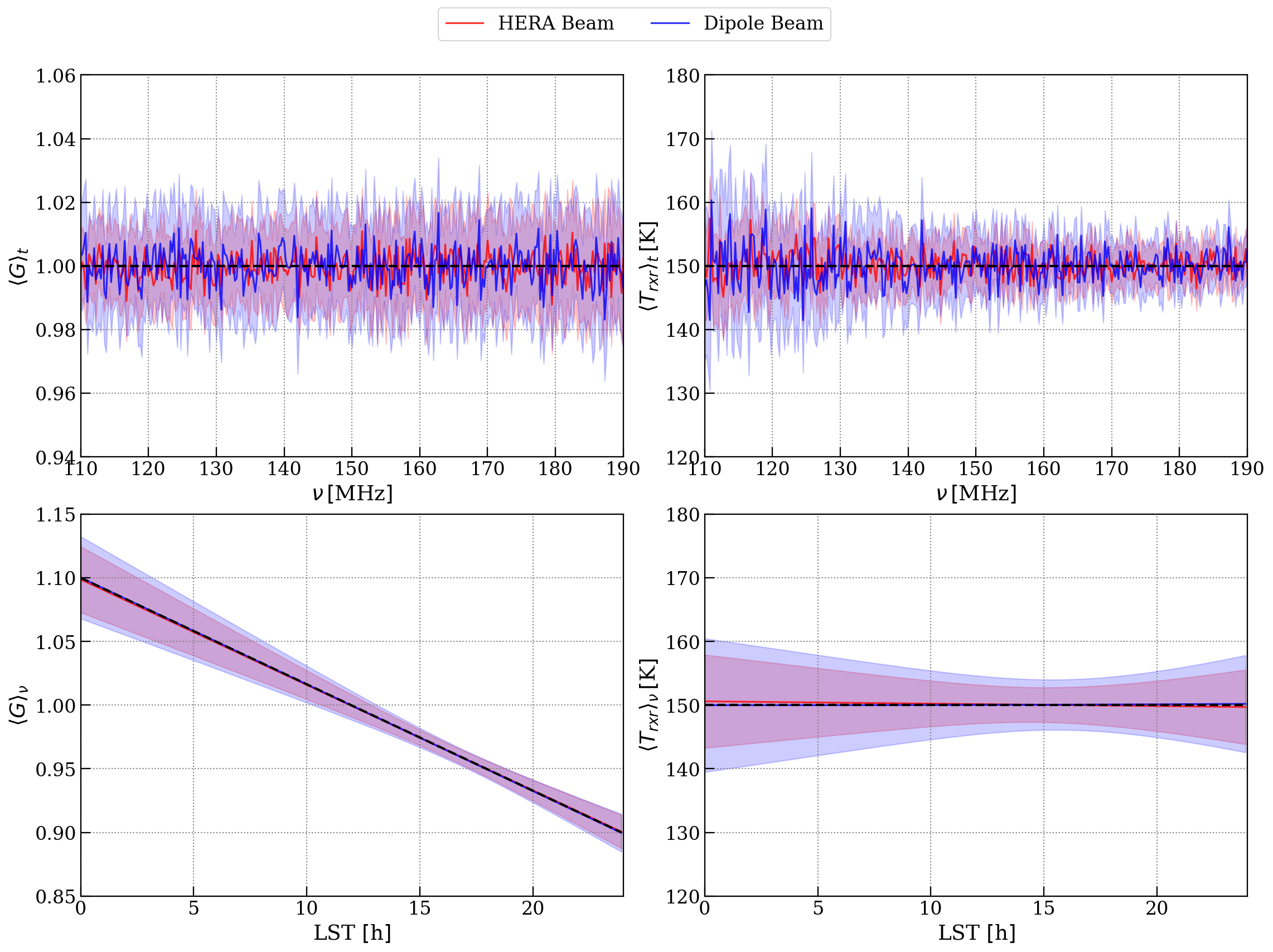}
    \caption{Calibration products for the modified calibration that uses first-order polynomials to fit for linear time-dependent gain $G(t)$ (see section~\ref{subsec:time_dependent_gain-model}). The top and bottom rows show calibration parameter $\hG$ (left column) and $\hT{rxr}$ (right column) averaged over time and frequency, respectively. Dashed lines show the input parameters averaged over corresponding axes. Shaded regions represent rms of error on the fit along the corresponding time and frequency axes used for averaging. Using a first-order polynomial in calibration mitigates the bias in calibration products due to temporal gain variations. However, corresponding uncertainty levels are increased relative to the reference case.} 
\label{fig:fit_G_t-dep-model}
\end{figure*}

\subsection{Time dependent gain model}\label{subsec:time_dependent_gain-model}

In section \ref{subsec:gain_variation}, we observed that in realistic cases where the receiver gain changes over time, fitting a constant gain term introduces significant bias in $\hG$ and $\hT{rxr}$. A viable method to mitigate this bias in calibration products is to incorporate gain time dependence in calibration and fitting for a polynomial gain term to account for temporal variation. In this section, we explore the addition of gain time dependence in the calibration method itself and fit for $\hG(t)$ and $\hT{rxr}(t)$. As a simple test case, we use a first-order polynomial to represent gain in calibration, i.e. a linear function of LST per channel. The modified fitting template becomes:

\begin{equation}\label{eqn:modified-template}
R[t] = (G_0 + G_1\times t)\times T_{\text{sky}}[t] + (n_0 + n_1\times t)
\end{equation}

where $G_0$ , $G_1$, $n_0$ and $n_1$ are fitting parameters. Calibration products $\hG$ and $\hT{rxr}$ can be obtained from these parameters as

\begin{equation}
    \begin{split}
        & \hat{G} = \hG_0 + \hG_1 \times t \, , \\
        & \hat{T}_{\text{rxr}} = \dfrac{\hG_0 + \hG_1 \times t}{\hat{n}_0 + \hat{n}_1 \times t} \, .
    \end{split}
\end{equation}

We use mock auto-correlations produced for case (a) in section \ref{subsec:gain_variation} for this analysis and fit for $\hG(t)$ and $\hT{rxr}(t)$. The above described fitting template should be able to capture the linear gain variation in mock auto-correlations perfectly. Figure \ref{fig:fit_G_t-dep-model} shows the corresponding calibration products averaged over time (top row) and frequency (bottom row). Note that the error bars on the averaged estimated parameters are the rms of the uncertainties from the fit along the frequency and time axes, respectively. We observe that bias in both $\hG$ and $\hT{rxr}$ is mitigated by incorporating a time-dependence of gain in calibration. However, the error on the fit for both antenna designs is larger compared to the reference case. As in the reference case, the calibration parameters for the dipole antenna show relatively larger uncertainty compared to the HERA antenna. The rms uncertainty on $\hG$ and $\hT{rxr}$ are $\text{rms}(G)\sim 0.7\%$, $\text{rms}(T)\sim 1.52\%$ for the HERA antenna and $\text{rms}(G)\sim 0.9\%$, $\text{rms}(T)\sim 2.2\%$ for the dipole antenna. The uncertainty on $\hT{rxr}$ for both antenna designs exhibits a frequency dependence, decreasing by a factor of about two between the lowest and the highest frequency; however it does not show any prominent spectral dependence in the case of $\hG$. This behavior is similar to that observed in the reference simulation, i.e., the spectral dependence of the sky affects the error on the additive term in the fit. We also notice that the fit is dominated by the LST range with the Galactic center above the horizon (12-24 hours) for which we see relatively small rms error compared to other LSTs.

In this analysis, we used a simple toy model to describe and mitigate the time dependence of antenna gain and the bias introduced in $\hG$ and $\hT{rxr}$ due to the same. However, in realistic cases, gain variation might be more sophisticated (e.g. sinusoidal), especially for observations covering the full LST and may not be approximated by a simple linear function. In such cases, the use of higher-order polynomials or other basis functions in calibration may be required (see e.g. \cite{wang2021} where Legendre polynomials are utilized similarly to calibrate the MeerKAT auto-correlation data). Additionally, prior understanding about instrumental gain drift and the LST dependence of ambient temperature may be used to account for temporal gain dependence in calibration. 

\begin{figure}
\centering
\includegraphics[width=\columnwidth]{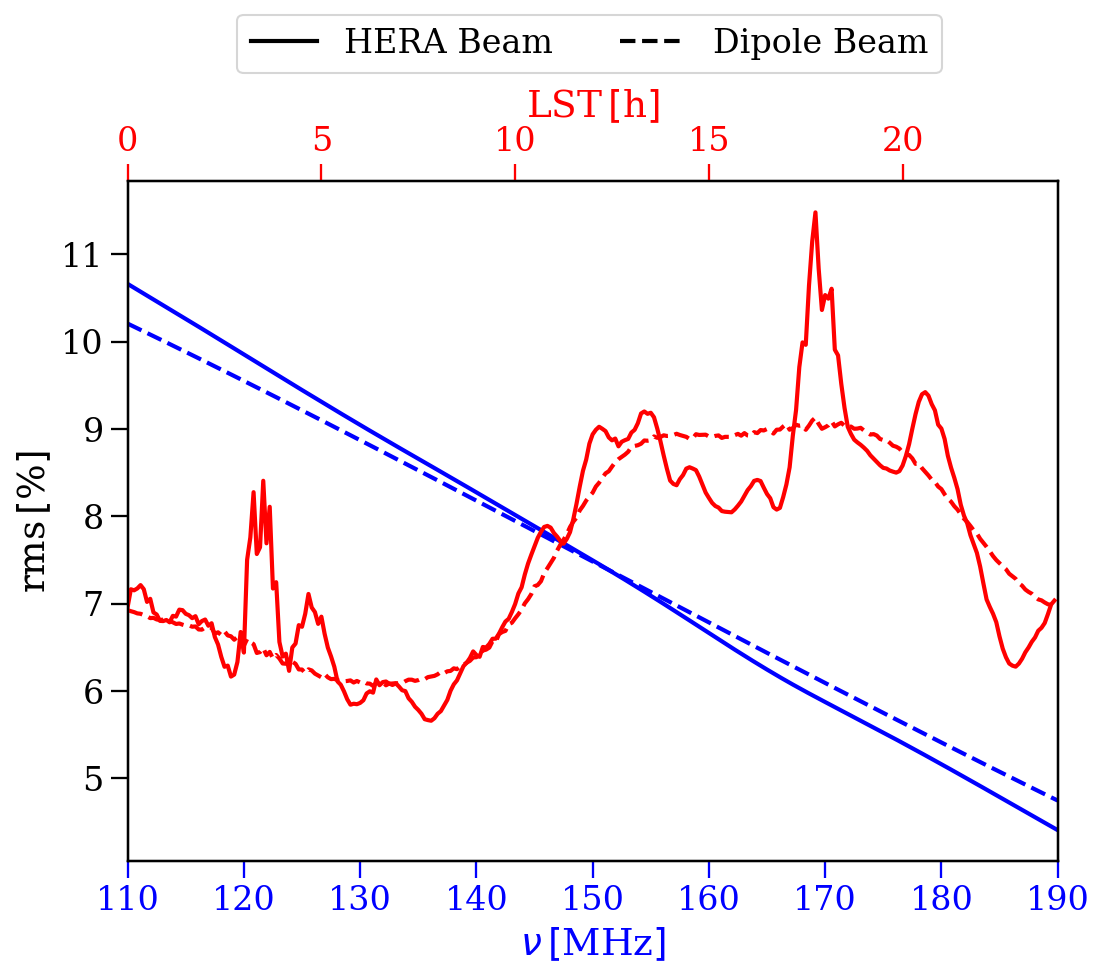}
    \caption{Comparison between beam averaged sky temperatures $\T{sky}$ obtained using GSM2008 and GSM2016 models for the two antenna designs. THe GSM2016 model shown here is the GSM2016 skymap at 200~MHz extrapolated to the required frequency range using a spectral index $\beta = -2.4$ unlike GSM2008 model produced for every frequency channel. Blue curves correspond to the rms of the fractional difference ($100\times(\T{sky}(\text{GSM2008})-\T{sky}(\text{GSM2016}))/\T{sky}(\text{GSM2008})$ taken over time axis for the two antenna designs. Red curves show the rms of the fractional difference taken over the frequency axis.} 
\label{fig:Tsky_GSM_comparison}
\end{figure}

\subsection{Effect of sky incompleteness}\label{subsec:sky-incompleteness}

Auto-correlations based calibration requires prior knowledge of sky brightness to obtain calibration products $\hG$ and $\hT{rxr}$. Sky models used in most interferometric calibration methods (either sky-based or using auto-correlations) are mainly based on radio sky surveys e.g. VLSS \citep{cohen2007}, GLEAM \citep{hurley-walker2016}, LOTSS\citep{shimwell2017,shimwell2019}, MSSS \citep{heald2015} carried out using various radio interferometers, as well as models and maps of diffuse Galactic emission e.g. Global Sky Models of diffuse emission \citep{deoliveiracosta2008,zheng2017}, LWA diffuse sky maps \citep{dowell2017,eastwood2018}. However, these surveys/maps will all, whether due to calibration, reconstruction, processing, or simply thermal noise, have inherent inaccuracy at some level. In the case of wide-field 21-cm arrays, sensitive in most cases to the entire visible sky, the most relevant limitation is an incomplete sky coverage or missing model flux at relevant size scale. When used for calibration, an incomplete sky model tends to introduce various errors in calibration products at different levels depending on the incompleteness. The impact of model incompleteness on interferometric calibration manifests as baseline dependent spectral structure which, unchecked, couples to all baselines during the calibration process \citep{patil2016,barry2016,ewall-wice2017}. In the auto-correlations, spatial structure emerges as time dependence, so it is worth investigating whether the spatial-spectral modulation leads to similar issues.

For this analysis, we assume that gains are stable in time and do not have a frequency structure (spectral structure due to the beam remains). Since sky model inaccuracy is a candidate for calibration bias, as a proxy for error, we use the older GSM model (GSM2008) \citep{deoliveiracosta2008} as the true sky and the newer GSM model (\cite{zheng2017}, GSM2016 hereafter) as the input sky model for calibration. The most notable difference between the two GSM models is that in the 2017 update, point sources have been removed. Point sources contribute a small fraction of the total power, emerging primarily as temporal variations on the scale of the beam crossing time; around half an hour for HERA, 3 hours for the dipole. However, the average spectrum of GSM2016 exhibits an artificial feature at 150 MHz (abrupt change in total amplitude) that may introduce unusual spectral behavior in calibration products. To avoid such complications, we extrapolate the GSM2016 map at 200 MHz to the desired frequency range using a spectral index $\beta = -2.4$. The spatial and spectral differences between GSM2008 and GSM2016 models provide a reasonable mismatch between the true sky and the sky model used for calibration. Figure \ref{fig:Tsky_GSM_comparison} shows rms of the fractional difference (in per cent) between the primary beam averaged sky temperature for the two models along time and frequency axis, respectively. Beam averaged sky temperatures for the two models differ from each other by approximately $5-12\%$ depending on LST and frequency. However, in the case of the HERA antenna, the spatial differences become more prominent due to its narrower field of view and appear as additional temporal structure compared to the dipole antenna with the wider field view that averages out finer spatial structures. Additionally, these variations are most prominent at LSTs when the Galactic center is above the horizon.

\begin{figure*}
\centering
\includegraphics[width=\textwidth]{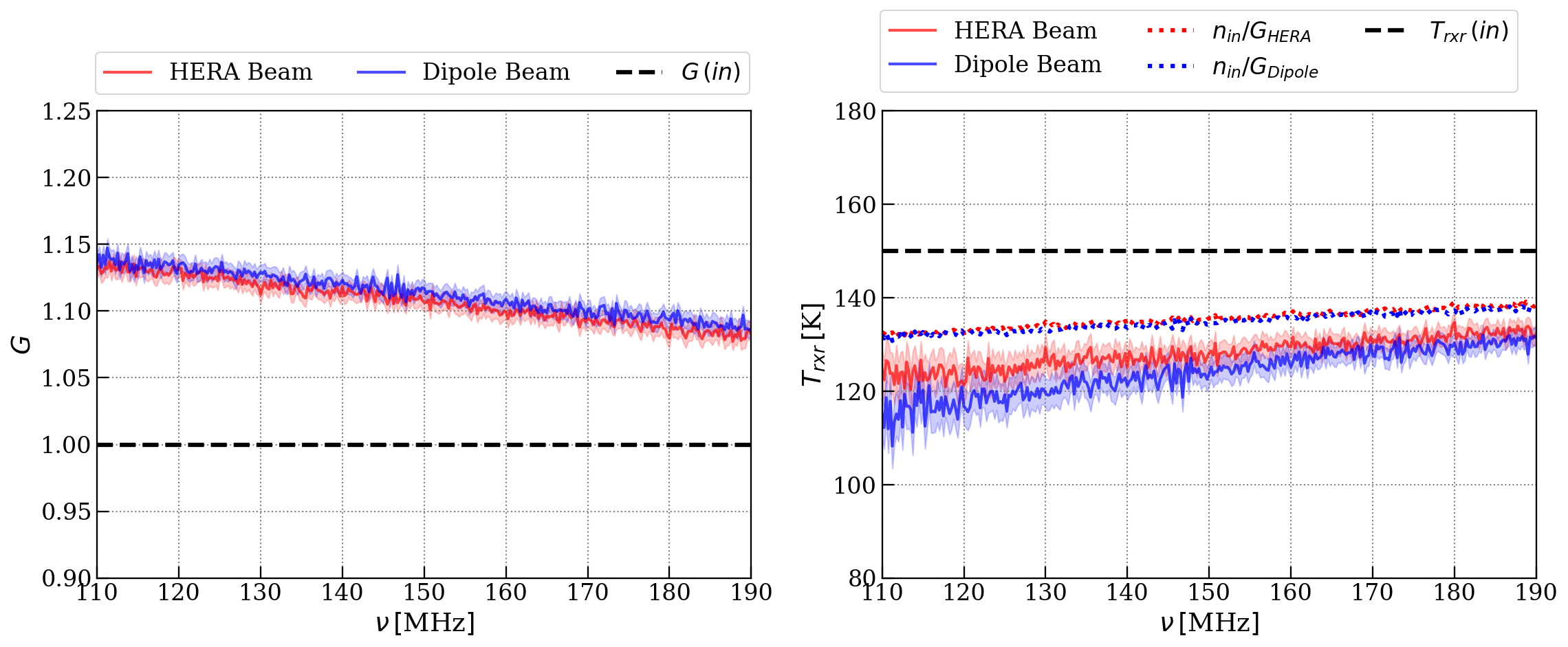}
    \caption{Fitting parameters for calibration using an incomplete sky model with incompleteness levels shown in figure~\ref{fig:Tsky_GSM_comparison}. Left panel: $\hG$ obtained from calibration for HERA and dipole primary beams. Right panel: same as the left panel but for $\T{rxr}$. Dotted curves show $n_{\text{in}}/\hG$, i.e. expected $\T{rxr}$ for the two antenna designs obtained from calibrated gains $\hG$ shown in the left panel and already known noise figure ($n_{\text{in}}$). Sky model incompleteness introduces spectrally varying bias in calibration products at similar levels for both antenna designs.} 
\label{fig:fit_sky-incompleteness}
\end{figure*}

Figure~\ref{fig:fit_sky-incompleteness} shows the calibration products when using an incomplete sky model for calibration. We observe that the incomplete sky model, when used for calibration, introduces bias in both $\hG$ and $\hT{rxr}$. For a given frequency channel, the amplitude of bias depends upon the level of sky-incompleteness, i.e. difference between the true sky and the sky model used for calibration. We also calculate the expected $\T{rxr}$ values (dotted curves in the right panel of figure~\ref{fig:fit_sky-incompleteness}) by dividing the already known noise figure (used to simulate corresponding auto-correlations) by $\hG$ obtained from calibration. We observe that these $\T{rxr}$ values are close to $\hT{rxr}$ values obtained from the calibration rather than the input $\T{rxr}$, suggesting that the sky-incompleteness used in the analysis affects $\hG$ more than the noise figure ($\hat{n}$) obtained from the calibration. In other words, the bias in $\hT{rxr}$ is mainly due to the bias in $\hG$ rather than $\hat{n}$. The rms error on $\hG$ and $\hT{rxr}$ have values $\text{rms}(G)\sim 0.3\%$ and $\text{rms}(T)\sim 1\%$ for both antenna designs, which is similar to error levels in the reference simulation. The sky model incompleteness at a certain level affects $\hG$ for both narrow and wide-field antenna designs the same way. However, the bias in $\hT{rxr}$ is relatively higher for the dipole beam at the lower frequency end compared to the HERA beam. Comparing figures~\ref{fig:Tsky_GSM_comparison} and~\ref{fig:fit_sky-incompleteness} we notice that the calibration of auto-correlations with an incomplete sky model introduces a spectrally varying bias in $\hG$ that is dependent on the sky model incompleteness level at corresponding frequencies. Therefore, it becomes evident that sky model incompleteness can play a crucial role in auto-correlation based interferometric calibration.       

\subsection{Effect of inaccurate primary beam}\label{subsec:beam-incompleteness}
The sky temperature model used for calibration, as described in section \ref{sec:autocalibration}, is the primary beam weighted average of the brightness temperature in every direction of the visible sky. The primary beam model used to calculate the average sky temperature needs to be accurate to obtain unbiased and accurate calibration products $\hG$ and $\hT{rxr}$. However, measurement of the primary beam for a given antenna is a daunting task by itself. Methods to estimate primary beams include electromagnetic simulations of radio antennas using software packages e.g. \textsc{cst} or \textsc{feko} \citep{fagnoni2021,mahesh2021}, holographic beam mapping techniques \citep{berger2016,iheanetu2019,asad2021}, beam mapping using bright radio sources transiting through the sky \citep{baars1977,nunhokee2020} as well as artificial sources e.g. satellites \citep{neben2016,line2018}, and more recently using artificial radio sources mounted on Unmanned Aerial Vehicles (e.g. drones) \citep{chang2015,jacobs2017}. Primary beam models obtained from these methods unavoidably contain small errors and thus differ (spatially and spectrally) from true antenna beams. Therefore, using these models may impact the calibration and possibly introduce bias in calibration. In this section, we study the effect of inaccurate primary beam models on auto-correlation based calibration.

\begin{figure*}
\centering
\includegraphics[width=\textwidth]{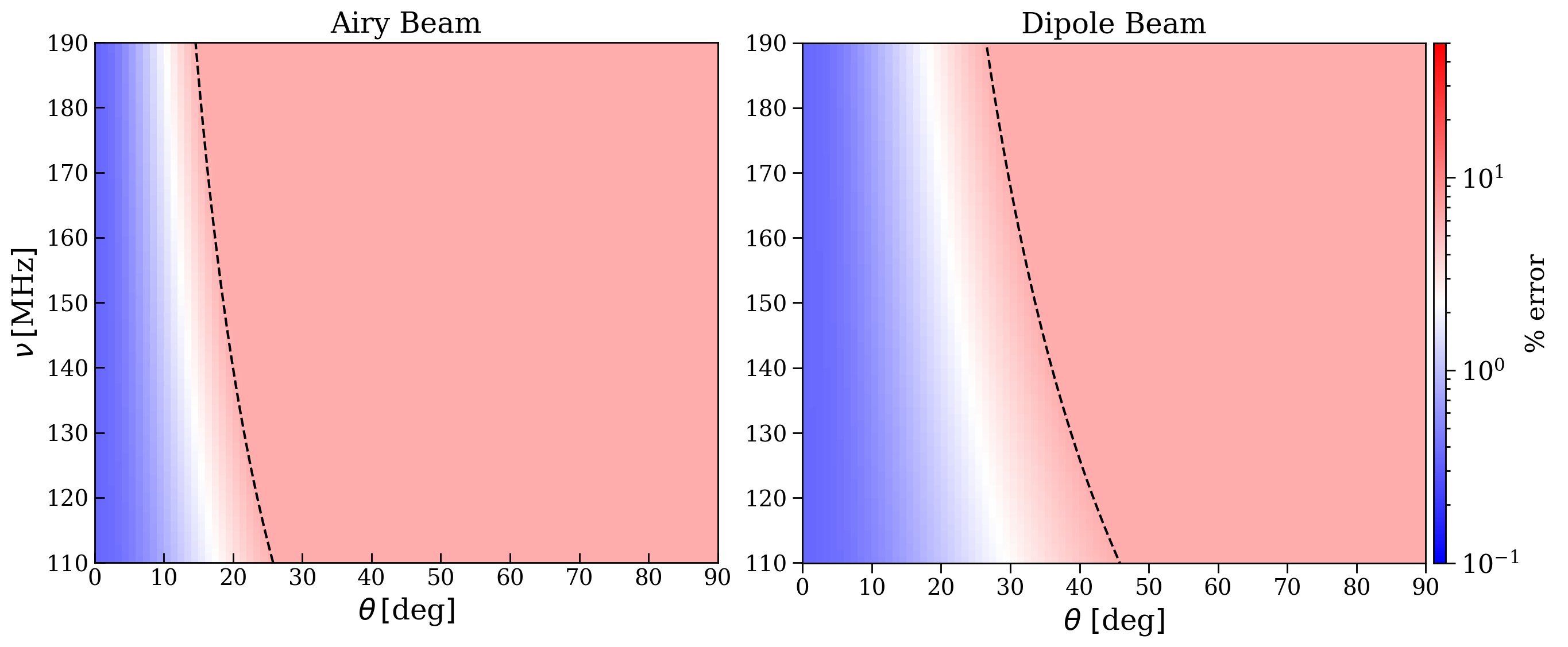}
    \caption{Fractional error profiles for the Airy beam (left panel) and the dipole beam (right panel). The dotted line corresponds to $\sin{^-1}(s_1)$ which represents the transition angle. Below this angle the fractional error is $<10\%$ while above this angle the error is equal to $10\%$.} 
\label{fig:beam_error}
\end{figure*}

For this analysis, we use an analytical Airy beam instead of the HERA beam used in previous sections to simplify the application of errors to the beam model. We define the analytical Airy beam model as

\begin{equation}
    A(\theta) = \left[ \dfrac{2J_1(\pi D \sin{\theta} / \lambda)}{\pi D \sin{\theta} / \lambda} \right]^2\,,
\end{equation}
where $D = 14$~m is the diameter of the HERA dish, and $J_1(x)$ is the Bessel function of first kind and order one. To represent primary beam inaccuracies for the Airy and dipole beams, we assume that primary beams are known with $<10\%$ errors within the main-lobe and have inaccuracies of $\approx 10\%$ for side-lobes, similar to the realistic errors reported in \cite{Neben2015} for MWA tile beam patterns. We use a similar approach as \cite{ewall-wice2017} to describe the beam errors. We describe the measured beam model (with errors) as
\begin{equation}\label{eqn:measured_beam}
    A_{E}(\theta,\phi) = A_0(\theta,\phi)\left[1 + f(\theta)\right]^2\,,
\end{equation}
where $A_{0}(\theta,\phi)$ is the true beam and $f(\theta)$ is the fractional error  applied to the true beam. $f(\theta)$ is written as
\begin{equation}\label{eqn:fractional_beam_error}
f(\theta) = \begin{cases} 
1 - (1-e_z)\exp{(-\sin^2{\theta}/2\sigma_e^2)} & |\sin{\theta}| < s_1 \\
1 - (1-e_z)\exp{(-s_1^2/2\sigma_e^2)} & |\sin{\theta}| \geq s_1 \\
\end{cases}
\end{equation}
where $e_z = 0.06$ and the parameter $s_1$ is given by
\begin{equation}
    \begin{split}
        & s_1^{\text{Airy}} =  \dfrac{7.0156\lambda}{\pi D}\,,\\
        & s_1^{\text{dipole}} = \sin \left( 0.5 \times \text{FWHM} \right)\,, \\
    \end{split}
\end{equation}
which is equivalent to the sine of the angular distance of the second null from the pointing centre in case of the Airy beam and sine of angular distance between the pointing centre and the half power point of the main-lobe (half of the FWHM) in case of the dipole beam. We define $\sigma_e = 1.5s_1$ to introduce frequency dependence in the fractional error. The parameters $e_z$, $s_1$ and $\sigma_e$ are set such that uncertainties approximately match the levels mentioned earlier. Fractional error profiles ($A_0(\theta,\nu)f^2(\theta,\nu)$) as a function of zenith angle and frequency for the two beams are shown in figure \ref{fig:beam_error}. 

\begin{figure*}
\centering
\includegraphics[width=\textwidth]{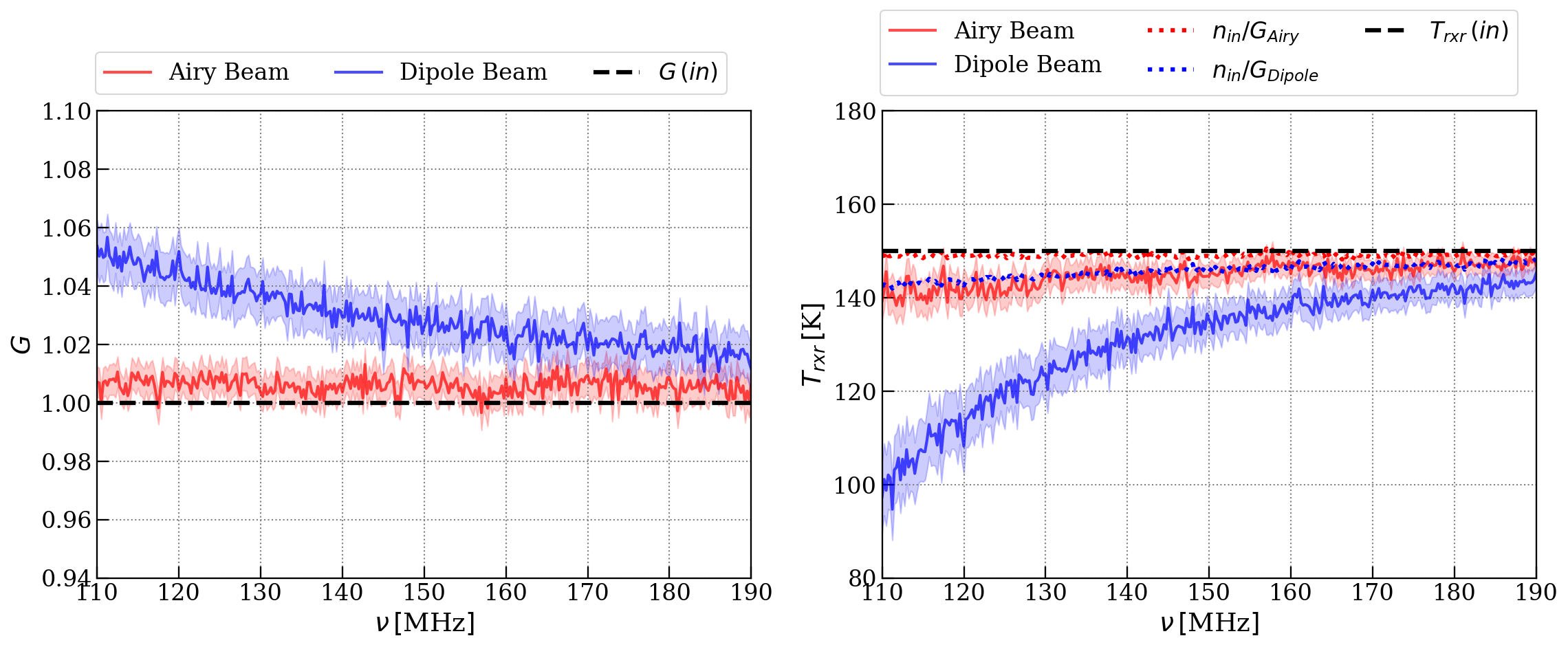}
    \caption{Calibration products for the calibration with an inaccurate beam model as described in section~\ref{subsec:beam-incompleteness}. Left panel: $\hG$ as a function of frequency for the Airy (red) and dipole (blue) beam. Right panel: same as the left panel but for $\hT{rxr}$. Dotted curves show $n_{\text{in}}/\hG$, i.e. expected $\T{rxr}$ for the two antenna designs obtained from calibrated gains $\hG$ shown in the left panel and already known noise figure ($n_{\text{in}}$). The dipole beam shows higher bias than the Airy beam.} 
\label{fig:fit_beam-incompleteness}
\end{figure*}

We simulate the mock auto-correlations using the ideal beam models without any error term ($A_0(\theta,\phi)$) whereas the sky models used as the input for calibration incorporate the measured beam model described by equation~\ref{eqn:measured_beam}. Calibration products as a function of frequency for this analysis are shown in figure \ref{fig:fit_beam-incompleteness}. We observe that the effect of primary beam inaccuracies is somewhat similar to that of sky model incompleteness as discussed in section \ref{subsec:sky-incompleteness}. An inaccurate beam model used for calibration seems to affect the dipole beam worse than the Airy beam. The bias in $\hG$ for the Airy beam is small enough that the estimated gain, for the most part, agrees with the input value, whereas $\hT{rxr}$ deviates from the input especially towards lower frequencies. On the other hands, calibration parameters for the dipole beam show bias with a steeper spectral dependence. We expect this behavior to be solely dependent on overall inaccuracy introduced in $\T{sky}$ due to the fractional error on the primary beam used for calibration. The expected $\T{rxr}$ values (dotted curves in figure~\ref{fig:fit_beam-incompleteness}) for both antenna designs are similar to the input $\T{rxr}$ suggesting that the bias in the noise figure $\hat{n}$ obtained from calibration dominates the bias in $\hT{rxr}$. The uncertainties on $\hG$ and $\hT{rxr}$ for the Airy beam are similar to the reference simulation, i.e. $\text{rms}(G)\sim 0.3\%$, $\text{rms}(T)\sim 1\%$, whereas the calibration parameters for the dipole beam show larger uncertainty with rms values of $\text{rms}(G)\sim 0.4\%$, $\text{rms}(T)\sim 1.5\%$. This analysis demonstrates that inaccurate beam models used in the calibration of auto-correlations introduce spectral structures in both gain and receiver temperature estimates with levels dependent upon the magnitude of beam inaccuracy and field of view of antenna elements. Additionally, primary beam inaccuracies seem to affect $\hat{n}$ more severely compared to $\hG$, making the former a leading cause of bias in $\hT{rxr}$.

\section{Delay spectrum analysis}\label{sec:delay-spec}
As we observed in previous sections, incompleteness effects such as temporal variation of instrumental gain, sky model incompleteness and primary beam inaccuracies introduce addition spectral structure in the calibration products $\hG$ and $\hT{rxr}$. When applied to the data, these calibration products may leak smooth foregrounds to otherwise clean modes in Fourier space. We use the delay transform technique to investigate this further. 

\begin{figure*}
\centering
\includegraphics[width=\textwidth]{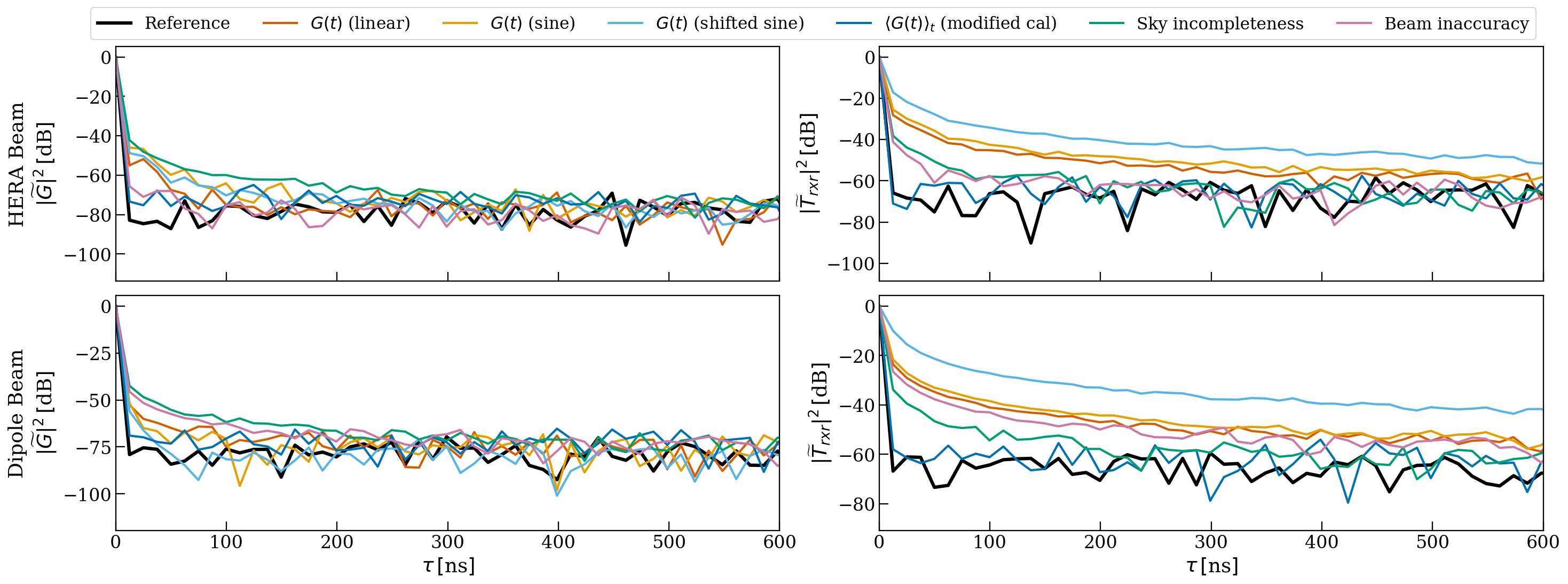}
    \caption{A delay space comparison of calibration products $\hG$ (left column) and $\hT{rxr}$ (right column) for different model incompleteness scenarios. The top and bottom rows correspond to HERA and dipole primary beams, respectively.} 
\label{fig:delayspec_combined}
\end{figure*}

Delay transform is a widely used diagnostic tool and a statistic in 21-cm experiments \citep{parsons2012,liu2014}. The delay spectrum is defined as the Fourier transform of a visibility spectrum observed by a given interferometric baseline along the frequency direction. The delay parameter $\tau$ is the Fourier dual to the frequency and corresponds to the time delay between the signal arriving (from a particular direction) at the two antennas of a given baseline. Although the delay spectrum is defined for a visibility, the methodology can also be applied to auto-correlations and other parameters such as gain and receiver temperature. A delay spectrum of any parameter with a frequency spectrum $F(\nu)$ is defined as

\begin{equation}\label{eqn:delay-spectrum}
    \Tilde{F}(\tau) = \int \text{e}^{2\pi i \nu \tau} F(\nu) d\nu \,.
\end{equation}

We compare the cases where model incompleteness incurs frequency-dependent bias in calibration products $\hG$ and/or $\hT{rxr}$ viz. temporal variation of antenna gain, sky incompleteness and primary beam inaccuracies, with the reference simulation. Additionally, we compare output products from the calibration method utilizing the time-dependent gain model (section~\ref{subsec:time_dependent_gain-model}) with the above-mentioned cases. Figure~\ref{fig:delayspec_combined} shows the delay space comparison of calibration products $\hG$ and $\hT{rxr}$ for different model incompleteness scenarios discussed in section~\ref{sec:model-incompleteness}. Note that the spectra are normalized with corresponding $\tau=0$ power. It essentially removes the mean bias in calibration products, and only the spectral structure remains. For both antenna designs (HERA/Airy beam and dipole beam), delay spectra of estimated gains ($\hG$) for temporal gain variations show additional power on small delay modes ($\tau<100$~ns) compared to the reference simulation. However, the additional power is not as prominent compared to the sky-incompleteness case. On the other hand, delay spectra of $\hT{rxr}$ show additional power on a wider range of delay modes especially in temporal gain variations. As discussed in section~\ref{subsec:time_dependent_gain-model}, we modified the calibration to fit for a first-order polynomial as a function of time to describe gain ($G$) and the noise figure ($n$). Here also, we observe that using a time-dependent model for gain and noise figure in calibration mitigates the additional power caused by temporal gain drifts in delay spectrum of $\langle\hT{rxr}\rangle_t$ on small delay modes  and the resulting delay spectra for both $\langle\hG\rangle_t$ and $\langle\hT{rxr}\rangle_t$ approximately match the reference simulation.

Sky model incompleteness also causes the power to leak to higher delays ($\tau < 200$~ns) in both $\hG$ and $\hT{rxr}$ delay spectra. Leakage of power to higher delays in $\hG$ due to sky incompleteness is strongest among other incompleteness scenarios. However, the effect is weaker in $\hT{rxr}$ compared to temporal gain variations. Finally, in the case of primary beam inaccuracy, the delay spectrum of $\hG$ for the Airy beam is closer to the reference simulation; however, $\hT{rxr}$ delay spectrum shows additional power on delay modes $\tau<200$~ns. The effect of beam inaccuracy, on the other hand, is stronger in the case of dipole beam and is similar to the sky-incompleteness effect for $\hG$. The $\hT{rxr}$ delay spectrum shows excess power on approximately all delay modes below $\tau \sim 600$~ns. This is expected as the dipole beam has a wider field of view than the Airy beam and is more sensitive to off-axis sky temperature. Therefore, uncertainties in the off-axis beam result in relatively large calibration errors and introduces additional spectral structure in calibration products.

\section{calibration of real data}\label{sec:real-data-calibration}
In this section, we apply the auto-correlation based calibration method on HERA and EDGES observation data to obtain corresponding $\hG$ and $\hT{rxr}$.

\begin{figure*}
\centering
\includegraphics[width=\textwidth]{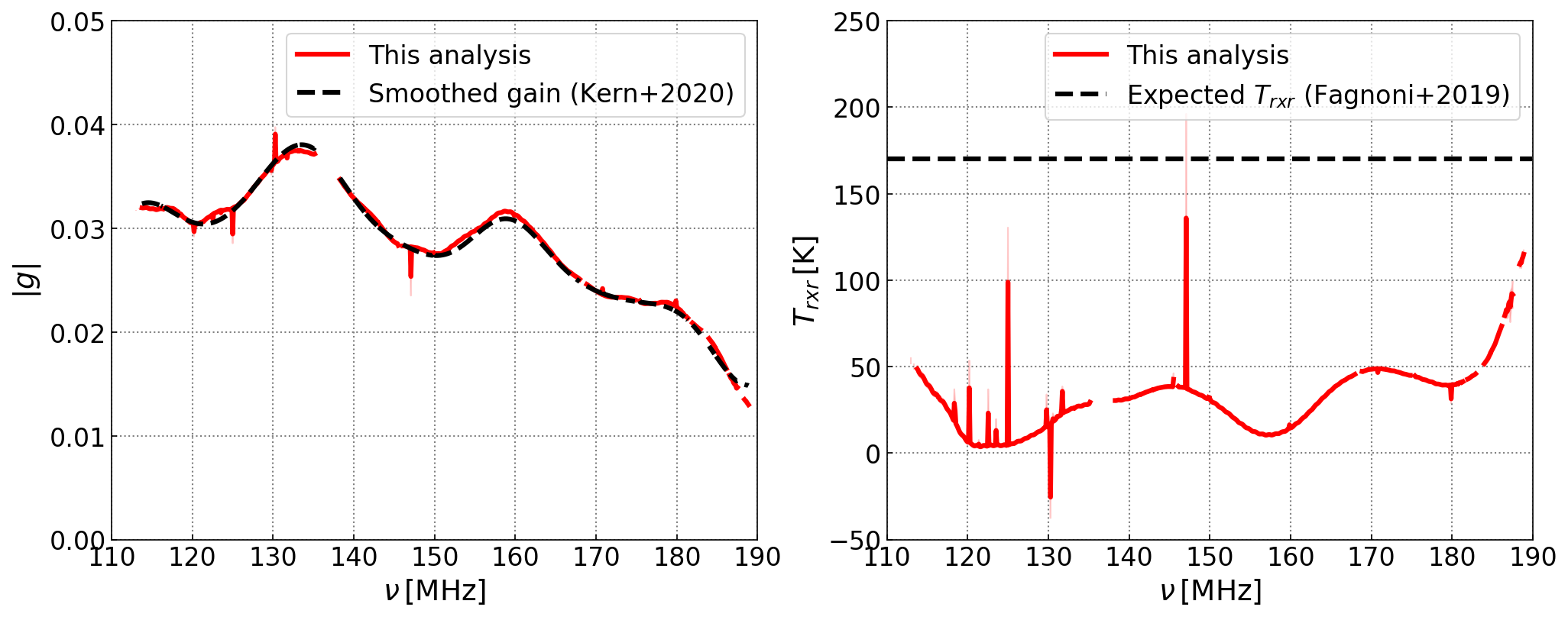}
    \caption{Calibration products obtained after calibrating the HERA auto-correlations with the method described in section~\ref{sec:autocalibration}. Red curves show $|g| = \sqrt{\hG}$ and $\hT{rxr}$ averaged over 8 antennas. The corresponding uncertainties (rms of the error on the fit over 8 antennas) are very small with levels $\text{rms}{|g|}<0.1\%$ and $\text{rms}{T}\sim 1\%$ and are not visible. The dashed curves in left and right panels correspond to the smoothed gain amplitude $|g|$ averaged over 8 antennas (reproduced from \citealt{kern2020b}) and the expected $\T{rxr}$ for HERA dishes \citep{fagnoni2021}, respectively.} 
\label{fig:HERA_Cal}
\end{figure*}

\subsection{HERA observations}

For this analysis, we use data from the HERA Phase-I (2017-18) observing cycle during which 47 HERA dishes (with PAPER type dipole feeds) were operational. We use the observation recorded on Julian Date (JD) 2458098 and select the 2-10~h LST range and 110-190~MHz frequency band for the analysis. The observation was recorded at time and frequency resolution of 10~s and 97.7~kHz. We further down-select auto-correlations for 8 antennas and single polarization (XX) from the observation data. To generate the sky model beam averaged sky temperature model for calibration, we use the GSM2008 sky model and the HERA dipole feed primary beam model from \cite{fagnoni2021}. We also account for the CMB temperature and the ground pickup (assuming a constant ground temperature of $300$~K) in the calculation of the sky temperature model to obtain a more realistic calibration model.

Figure~\ref{fig:HERA_Cal} provides a comparison of gain amplitude $|g| = \sqrt{\hG}$ obtained using the method presented here with the gain amplitude obtained by calibrating cross-correlation visibilities as described in \cite{kern2020a}, and a comparison of $\hT{rxr}$ obtained with expected value of $\T{rxr}\sim 170$~K for the HERA dish with dipole feed \citep{fagnoni2021}. We observe that the average gain amplitude obtained from auto-correlations matches very well with the frequency smoothed gain amplitude obtained from cross-correlation visibilities of the same dataset. On the other hand, $\hT{rxr}$ is underestimated with a significant bias level compared to the expected value of $\T{rxr}$. We assume the data covariance equal to identity i.e. $\textbf{C} = \textbf{I}$ when determining the calibration parameters. This results in very small fit uncertainties on both $\hG$ and $\hT{rxr}$ with levels $\sigma_{g} \lesssim 0.1\%$ and $\sigma_{T} \sim 1\%$. We also calibrated the data using the modified calibration (discussed in section~\ref{subsec:time_dependent_gain-model}) to account for any temporal gain that may be present in the data, however it does not impact $|g|$ and slightly improves the $\hT{rxr}$ estimate but only by a few per cent (plot not shown here). Presence of the spectral structure at $\sim30$~MHz level in $\hT{rxr}$ seems to suggest frequency-dependent inaccuracies in the beam model to be the main cause of the bias in $\hT{rxr}$. Even though this behavior is similar to the biases observed in the primary beam inaccuracies simulation discussed in section~\ref{subsec:beam-incompleteness}, the  inaccuracy level used in the simulation does not produce a significantly high bias in $\hT{rxr}$ compared to the calibration of HERA auto-correlations. Therefore, we suspect that other factors that remain unaccounted for, such as cable reflections, mutual coupling and ground temperature model also contribute to the bias in $\hT{rxr}$ in addition to the primary beam inaccuracies. We expect that incorporating improved beam models (that include mutual coupling and finer frequency sampling) and accounting for other factors in future analyses will mitigate the bias in $\hT{rxr}$ estimates.

\begin{figure*}
\centering
\includegraphics[width=\textwidth]{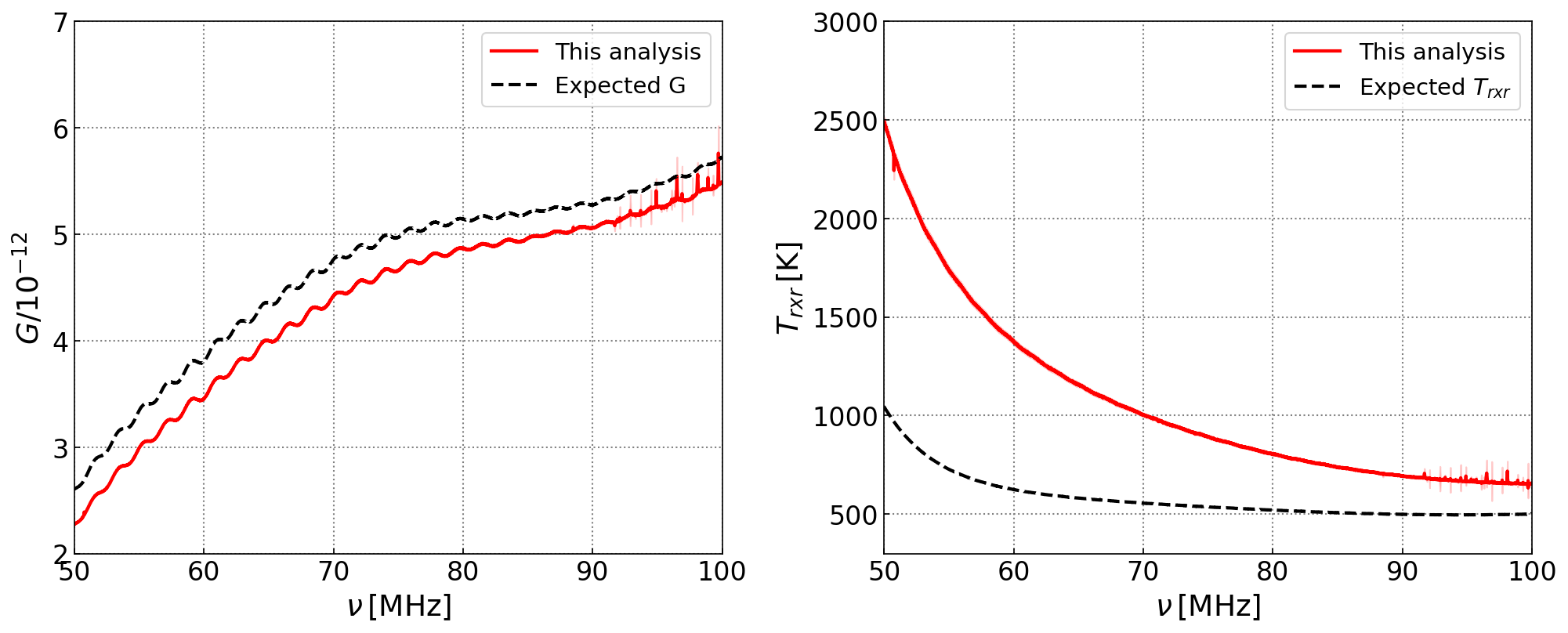}
    \caption{$\hG$ (left panel) and $\hT{rxr}$ (right panel) obtained after calibrating EDGES data. Dashed black curves correspond to expected $G$ and $\T{rxr}$ values obtained by propagating already known 3-position switching calibration parameters of EDGES2 receiver system for the observation data used in the analysis.} 
\label{fig:EDGES_Cal}
\end{figure*}

\begin{table*}
\centering
\caption{A summary of simulation input, calibration output for various model incompleteness effects, error and bias levels for various incompleteness effects investigated in the analysis.}

\label{tab:summary_effects}
\begin{tabular}{p{3.5cm}x{4cm}x{4cm}p{4cm}}
\hline
\textbf{Incompleteness effect} & \textbf{Simulation input} & \textbf{Calibration output} & \textbf{Comments} \\
\hline
Reference simulation & Constant $G$, fixed $\T{rxr}$, Perfect sky model and beam & $\hG$ and $\hT{rxr}$ without bias and small error & Ideal scenario\\
\hline
Temporal gain variation: linear, sine, shifted sine & $G(t)$, fixed $\T{rxr}$, Perfect sky model and beam & $\hG$ and $\hT{rxr}$ with bias & errors in calibration products larger than the reference simulation\\
\hline
Modified calibration with time dependent gain model & $G(t)$, fixed $\T{rxr}$, Perfect sky model and beam & $\langle \hG(t)\rangle_t$ with no observed bias, $\langle \hT{rxr}(t)\rangle_t$ with no observed bias & rms error (along time) for $\hG(t)$ and $\hT{rxr}(t)$ is larger than the reference.\\
\hline
Sky incompleteness & Constant $G$, fixed $\T{rxr}$, incomplete sky model, perfect beam & $\hG$ and $\hT{rxr}$ with frequency dependent bias & errors in calibration products similar to the reference for HERA beam and slightly larger for dipole beam. \\ 
\hline
Primary beam inaccuracy & Constant $G$, fixed $\T{rxr}$, perfect sky, inaccurate beam & $\hG$ and $\hT{rxr}$ with frequency dependent bias & errors larger than the reference simulation, Airy beam shows small errorr/bias than the dipole beam.\\
\hline
\end{tabular}
\end{table*}

\subsection{EDGES observations}

We also use total power data observed with the EDGES2 instrument for the auto-correlation based calibration. The data was observed in 2016 (Day:260) and used in the analysis in \cite{bowman2018}. Note that we utilize the raw data (uncalibrated) for this analysis. The observation spans the full 24 hours LST range and has a time and frequency resolution of 39~s and 6.1~kHz. We down-select the frequency range of the data to 50-100~MHz. For the calculation of beam averaged sky temperature model ($\T{sky}(\nu,t)$), we use the Haslam 408~MHz sky-map of diffuse emission \citep{haslam1982} extrapolated to the frequency range of the EDGES data using a spectral index value $\beta = -2.55$. We also correct for the CMB temperature and the ground pickup to obtain a more accurate model. The primary beam model for average sky temperature calculation is taken from \cite{mahesh2021} who simulated the EDGES primary beam using the \textsc{feko} software package. 

The resulting $\hG$ and $\hT{rxr}$ obtained from calibration are shown in figure~\ref{fig:EDGES_Cal}. The expected gain and receiver temperature values are calculated by propagating already known 3-position switching calibration parameters (noise wave parameters and reflection coefficients) for the above observation to obtain equivalent gain and receiver temperature (see e.g. \citealt{rogers2012,monsalve2017}). We find that the gain amplitude $\hG$ is underestimated compared to the expected value. On the other hand, the $\hT{rxr}$ is highly overestimated compared to the expected value. Bias in $\hG$ and $\hT{rxr}$ have slow frequency dependence with levels varying between $4-12\%$ and $30-140\%$, respectively. Similar to the HERA calibration, we assume identity data covariance here which leads to uncertainties on $\hG$ and $\hT{rxr}$ between $0.08 - 0.1\%$ and $0.25-0.5\%$ respectively, with higher error at lower frequency end. 

We further investigate the cause of the bias in $\hG$ and $\hT{rxr}$ by changing the sky model used for the calibration of the EDGES data. We repeated the calibration for using the following sky models: extrapolated Haslam map with spectral indices $\beta = -2.5, -2.55, -2.6$, and $-2.65$ (without CMB temperature correction), and extrapolated GSM2008 map (at 200~MHz) with $\beta=-2.55$. We find that $\hT{rxr}$ is still overestimated regardless of the choice of the sky model (plots not shown here). However, the overall bias in both $\hG$ and $\hT{rxr}$ changes with spectral index and the calibration favors the Haslam sky model with a shallower spectral index of $\beta=-2.5$. Changing the spectral index does not affect the noise figure estimate ($\hat{n}$) but results in a different $\hG$ estimate(and hence $\hT{rxr}$). Furthermore, changing the sky map from Haslam to GSM2008 or GSM2017 for a given spectral index ever so slightly affects the shape of $\hG$ and $\hT{rxr}$ along frequency but overall bias levels are approximately similar for both cases. In summary, error in fitted noise figure $\hat{n}$ is smooth in frequency and insensitive to changes in the sky spectral index but it is sensitive to changes in spatial structure. In contrast, $\hG$ is affected by both spectral index and the spatial structure of the sky model. The bias in $\hG$ and the frequency smoothness of $\hT{rxr}$ (similar behavior observed in simulation discussed in section~\ref{subsec:sky-incompleteness}) indicates the sky incompleteness to be the primary cause of the discrepancy between the expected and calibrated values of $G$ and $\T{rxr}$, however, the level of the sky incompleteness used in simulation does not fully explain the bias in $\hT{rxr}$.


\section{Summary and Discussion}\label{sec:summary}

Low-frequency 21-cm cosmology experiments aiming to measure the redshifted 21-cm signal from the Cosmic Dawn and Epoch of Reionization require extremely accurate and precise calibration of instruments to extract the faint 21-cm signal from the observed data. However, calibration methods are susceptible to biases and uncertainties due to incompleteness in calibration model e.g. unstable instrumental gains (in time and frequency), incomplete sky model, or primary beam model inaccuracies. Calibration of bandpass amplitude and receiver temperature of instruments using auto-correlations (or total power measurements) against known primary beam and sky-brightness, explored by several 21-cm experiments, is also susceptible to uncertainties and biases due to model incompleteness. We used simulations to investigate various effects that impact the auto-correlation (or total power) based calibration. Our findings are enumerated in the following list and further summarized in table \ref{tab:summary_effects}.

\begin{enumerate}
    
    \item Temporal variation in antenna gains introduces a bias in estimated gain and receiver temperature with respect to the time invariant simulation input. The bias in receiver temperature shows a slow spectral dependence. Calibration products for the HERA dish show a weak spectral ripple for all three gain variation scenarios, due to leakage from the HERA primary beam into the passband. The bias is relatively higher for the dipole antenna design, possibly due to larger average uncertainty resulting from its wider field of view.
    
    \item We show that it is possible to mitigate the bias in calibration products incurred due to temporal variation. This requires modifying the calibration step to solve for a time dependent gain and the noise figure rather than solving for a single gain and noise figure value per frequency channel. We show that using a first-order polynomial model for gain and noise figure in calibration mitigates the bias incurred due to linear variation of antenna gain. However, the modified calibration increases the rms noise levels in recovered gain and receiver temperature by a factor of a few.
    
    \item Using an incomplete sky model for calibration introduces spectral structure in the estimated gain and receiver temperature that depends on the level of incompleteness. Both antenna designs are similarly impacted by this effect.

    \item Inaccuracies in primary beam model used in calibration also lead to frequency-dependent bias in the estimated gain and receiver temperature. However, the bias is significantly smaller for the Airy beam than the dipole beam which has a wider field of view. This is mainly due to lower off-axis sensitivity of the Airy beam compared to a dipole beam. This, combined with off-axis errors in the dipole primary beam results in larger bias. 
    
    \item The delay spectrum analysis confirms the leakage of power to non-zero delay modes suggesting that various model-incompleteness effects introduce spectral structure to calibration products.
    
    \item We also show the application of auto-correlation based calibration on real HERA and EDGES data. The gain amplitudes obtained for HERA auto-correlations match well with the gain amplitudes obtained from calibration of HERA cross-correlation visibilities. However, the receiver temperatures are severely underestimated compared to expected values. Although we suspect this to be caused by primary beam inaccuracies as shown by the simulation, the inaccuracy level used in simulation does not explain bias level in the HERA receiver temperature. Possible contribution from other unaccounted factors such as mutual coupling, cable reflections, ground temperature model may explain the unexpected bias level in HERA receiver temperature. 
    
    \item Calibration products for EDGES data disagree with the EDGES internal calibration but in the opposite sense from HERA. In the case of EDGES, the noise figure and hence the receiver temperature is overestimated. The fitted noise figure varies most strongly when changing between different spatial models (e.g. Haslam vs GSM2008) and not at all when changing the spectral index for a given sky model. This suggest the spatial accuracy of the sky catalog to be more relevant than the spectral index. We left the exploration of beam model variations and their effect on EDGES and HERA data calibration for analyses.
   
\end{enumerate}

In summary, we find that auto-correlation based calibration is sensitive to model inaccuracies like  temporal gain variations, sky incompleteness and primary beam inaccuracies. Applying the calibration method to data from HERA and EDGES we find small biases in gain and larger unexpected offsets in receiver temperature. The behavior of these offsets are similar to those found when injecting model error into simulation, however the latter does not completely explain the scale of the offsets in receiver temperatures obtained from calibration of both instruments. We show that it is possible to mitigate bias caused by temporal gain variations by using higher-order temporal polynomials to represent the calibration products. Mitigation of errors due to sky and beam inaccuracies will improve with refined models of those factors.

\section*{Acknowledgements}
This work was supported by the National Science Foundation through awards for HERA (AST-1836019) and EDGES (AST-1813850, AST-1609450, and AST-1908933). NM was supported by the Future Investigators in NASA Earth and Space Science and Technology (FINESST) cooperative agreement 80NSSC19K1413. DL was supported by the NASA Solar System Exploration Virtual Institute Cooperative Agreement number 80ARC017M0006. N. Kern gratefully acknowledges support from the MIT Pappalardo Fellowship. HERA is hosted by the South African Radio Astronomy Observatory, which is a facility of the National Research Foundation, an agency of the Department of Science and Innovation. EDGES is located at the Murchison Radio-astronomy Observatory. We acknowledge the Wajarri Yamatji people as the traditional owners of the Observatory site. We thank CSIRO for providing site infrastructure and support. This analysis makes use of following software packages (publicly-available and open-source): \textsc{pygdsm}(\url{}https://github.com/telegraphic/pygdsm), \textsc{healpy} (\url{https://pypi.org/project/healpy/}), \textsc{ephem} (\url{https://pypi.org/project/ephem/}), \textsc{numpy} (\url{https://numpy.org/}), \textsc{scipy} (\url{https://www.scipy.org/}), and \textsc{matplotlib} (\url{https://matplotlib.org/}).

\section*{Data availability}
The data underlying this article will be shared on reasonable request to the corresponding author.



\bibliographystyle{mnras}
\bibliography{bibliography.bib} 


\bsp	
\label{lastpage}
\end{document}